\documentclass[twocolumn]{aastex62}

\usepackage{amsmath}
\usepackage[utf8]{inputenc}

\defcitealias{kessler19}{K19}
\defcitealias{lochner16}{L16}
\defcitealias{revsbech18}{R18}
\newcommand\plasticcresults{Hlo\v{z}ek et al. (2019, in prep.)}

\graphicspath{{./}{figures/}}

\newcommand{\changeda}[1]{#1}
\newcommand{\changedb}[1]{#1}
\newcommand{\changedc}[1]{#1}
\newcommand{\changedd}[1]{#1}

\shorttitle{Gaussian Process Augmentation for Photometric Classification}
\shortauthors{Boone}

\begin{document}

\title{Avocado: Photometric Classification of Astronomical Transients \\ with Gaussian Process Augmentation}

\correspondingauthor{Kyle Boone}
\email{kboone@berkeley.edu}

\author[0000-0002-5828-6211]{Kyle Boone}
\affiliation{Physics Division, Lawrence Berkeley National Laboratory, 1 Cyclotron Road, Berkeley, CA, 94720, USA}
\affiliation{Department of Physics, University of California Berkeley, 366 LeConte Hall MC 7300, Berkeley, CA, 94720-7300, USA}





\begin{abstract}

Upcoming astronomical surveys such as the Large Synoptic Survey Telescope (LSST) will rely
on photometric classification to identify the majority of the transients and variables that they discover.
We present a set of techniques \changeda{for photometric classification that can be applied} even when the
training set of spectroscopically-confirmed objects is heavily biased towards
bright, low-redshift objects. \changeda{Using Gaussian process regression to model arbitrary light curves
in all bands simultaneously, we ``augment'' the training set} by generating
new versions of the original light curves covering a range of redshifts and observing conditions. We train a
boosted decision tree classifier on features extracted from the augmented light curves, and we show how such
a classifier can be designed to produce classifications that are independent of the redshift distributions of objects 
in the training sample. Our classification algorithm was the best-performing among the 1,094 models considered in
the blinded phase of the Photometric LSST Astronomical Time-Series Classification Challenge (PLAsTiCC),
scoring 0.468 on the \changeda{organizers' logarithmic-loss metric} with flat weights for all object
classes in the training set, and achieving an AUC of 0.957 for classification of Type Ia supernovae.
\changedd{Our results suggest that spectroscopic campaigns used for training photometric classifiers
should focus on typing large numbers of well-observed, intermediate redshift transients instead of attempting
to type a sample of transients that is directly representative of the full dataset being classified.} All
of the algorithms described in this paper are implemented in
the \href{https://www.github.com/kboone/avocado}{\texttt{avocado} software
package}\footnote{\url{https://www.github.com/kboone/avocado}}.

\end{abstract}



\section{Introduction} \label{sec:introduction}

Upcoming large-scale optical astronomical surveys will collect images for most of the visible
sky on a nightly to weekly basis, discovering large numbers of astronomical
transients and variables every night. Classifying these objects is essential to
perform further scientific analyses on them. Traditionally, transients and variables are classified
using spectroscopic followup. However, spectroscopic resources are limited, so future surveys will
have to rely heavily on photometric classification methods.

One major application of photometric classification is cosmological measurements with Type Ia
supernovae (SNe~Ia). Distance measurements with SNe~Ia led to the initial discovery of \changeda{the accelerating
expansion of the universe} \citep{riess98, perlmutter99}. Subsequent studies have collected a sample of over 1,000
spectroscopically-confirmed SNe~Ia, providing increasingly strong constraints on the
properties of dark energy \citep{knop03, riess04, astier06, kowalski08, suzuki12, betoule14, scolnic18}.
With modern surveys, the discovery rate of SNe~Ia is rapidly
outpacing the growth of resources to acquire spectroscopic classifications. The Dark Energy
Survey \citep[DES,][]{des05} was projected to acquire spectroscopic classifications for only 20\% of their
sample of up to 4,000 SN~Ia light curves \citep{bernstein12}. Similarly, the
Pan-STARRS Medium Deep Survey \citep[PS1,][]{kaiser10} discovered over 5,000 likely supernovae,
but only obtained spectroscopic classifications for 10\% of this sample \citep{jones17}.
\changeda{Upcoming large-scale} surveys
such as the Large Synoptic Survey Telescope \changedc{\citep[LSST,][]{lsst09}} are projected to obtain light curves for
$\sim$100,000 SNe~Ia \citep{desc18}, and will almost certainly have spectroscopic classifications
for a much smaller fraction of their full sample.

Cosmological analyses with photometrically-classified SNe~Ia are complicated by
the fact that there is contamination from other transients in the sample, such as Type~Ib/c or Type~II
supernovae. These other transients do not have the same intrinsic luminosity as SNe~Ia,
and they will bias cosmological measurements if they are \changeda{accidentally} included in a cosmological analysis.
In principle, unbiased cosmological parameters can be recovered from photometrically-classified samples of SNe~Ia
by using Bayesian methods to model the contamination of the non-SN~Ia transients
in the sample \citep{kunz07, hlozek12, rubin15, jones17}. The performance of these methods
depends heavily on their ability to distinguish SNe~Ia from other transients, so accurate
photometric classifiers are essential to their operation.

There are several major challenges to designing a photometric classification algorithm.
The light curves generated by surveys such as LSST are sparsely sampled, and the observations do
not occur on regular time intervals. Observations occur in different bands, and only a subset of
the bands are typically available on any given night. The uncertainties on the photometry are
heteroskedastic, and some bands have much higher noise levels than others. \changeda{Currently,} photometric
classifiers are typically trained on a subset of light curves from the survey in question that have
spectroscopic confirmation. Brighter, nearby transients are
significantly easier to spectroscopically classify than fainter, more distant ones, so
training sets for transient surveys will typically be highly biased towards bright, nearby objects. 

To understand the performance and limitations of photometric classifiers for DES, the Supernova Photometric
Classification Challenge \citep[SNPhotCC,][]{kessler10} was initiated. The organizers of this challenge
produced a simulation of Type~Ia, Ib, Ic and II supernovae observed with realistic DES observing
conditions. The SNPhotCC dataset consists of a training set of 1,103 spectroscopically confirmed objects
and a test set of 20,216 objects without spectroscopic confirmation. Participants were challenged to
develop classifiers that could use the known labels of the training set to infer the types of objects in
the test set.

A wide variety of models and techniques were developed for, or applied to, data from the SNPhotCC.
The techniques that have been applied to photometric classification on this dataset include Bayesian template
comparisons \citep{poznanski07, sako11}, diffusion maps with random forests \citep{richards12},
neural networks \citep{karpenka13}, kernel PCA with nearest neighbours \citep{ishida13}, convolutional neural
networks \citep{pasquet19}, and deep recurrent neural networks \citep{charnock17}.
\citet{lochner16} (hereafter: \citetalias{lochner16}) compared the performance of several different machine
learning algorithms on the SNPhotCC dataset, and found that fitting the SALT2 model of SNe~Ia
\citep{guy07} to observations and training a boosted decision tree on the parameters of that model \changeda{gave}
the best classifier performance of the methods that they tested. 

The major concern with all of these photometric classification methods is that they have poor performance
when the training set of objects with spectroscopically-determined types is not representative of the
full dataset. \citetalias{lochner16} achieve an Area under the Receiver Operator Characteristic Curve
(AUC, defined in Section~\ref{sec:single_class_metrics}) of 0.98 when the training set is representative of the full dataset,
but an AUC of only 0.88 when training on the non-representative training set in the SNPhotCC. \citet{revsbech18}
(hereafter: \citetalias{revsbech18}) introduced the first effective attempt to deal with non-representative
training sets in a model that they call \texttt{STACCATO}. They augment the original training data by generating new light curves from
ones in the training sample to produce a new training set that is more representative of the full dataset.
\texttt{STACCATO} achieves an AUC of 0.96 when trained on their augmented training set compared to 0.92 when trained
on the original training set.

\changedc{The majority} of the previously discussed classifiers were trained and evaluated on the SNPhotCC dataset. Following
the success of the SNPhotCC, a new challenge was created focusing on photometric classification
for the LSST. This challenge, the Photometric LSST Astronomical Time-Series Classification Challenge
\citep[PLAsTiCC,][hereafter: \citetalias{kessler19}]{kessler19}, \changeda{includes 18 different kinds of transients
and variables, and is not limited to different kinds of supernovae like the SNPhotCC was.} From September~28, 2018 to
December~17, 2018, a blinded version of the PLAsTiCC dataset was provided through the
Kaggle platform\footnote{\url{https://www.kaggle.com/c/PLAsTiCC-2018}} with class labels available
only for the training set of spectroscopically-classified objects. Teams were challenged to determine the
types for the remainder of the dataset, and submit their predictions to the Kaggle platform where
a score was assigned to their predictions. A total of 1,094 teams submitted predictions as part of
this challenge. We developed a new classifier for the PLAsTiCC dataset that expands on the previously described
techniques. Out of all of the models submitted during the blinded phase of the PLAsTiCC, our classifier
achieved the best performance on the PLAsTiCC test set measured using the weighted log-loss metric proposed
by the PLAsTiCC team \citep{malz18}.

In this paper we discuss several new \changeda{techniques} that we developed to improve the performance of photometric
classifiers when trained on a spectroscopically-confirmed light curve sample that is not representative of
the full light curve sample. We show how a Gaussian process in both time and wavelength can be used to build
a light curve model that can incorporate information from all bands simultaneously. Using such a model, we can
then take a training set that is heavily biased towards bright, low-redshift objects, and ``augment'' it by
generating new light curves covering a range of redshifts and observing conditions. This augmented dataset
contains light curves that are much more representative of the full light curve sample. We then show
how a classifier can be trained whose performance is independent of the redshift distributions of the different
object types in the training sample.

In Section~\ref{sec:dataset}, we discuss the PLAsTiCC dataset along with
several metrics that can be used to evaluate the performance of photometric classifiers on this dataset. We
describe the new \changeda{techniques} that we developed for photometric classification in Section~\ref{sec:methods}.
In Section~\ref{sec:results} we discuss the performance and limitations of our classification
\changeda{techniques}, and compare them to other techniques. Finally, in Section~\ref{sec:discussion}, we discuss future
work that could be done to improve classifier performance, and how the techniques described in this paper
could be applied to other classifiers.

\section{Dataset} \label{sec:dataset}

\subsection{The PLAsTiCC dataset} \label{sec:plasticcdataset}

The PLAsTiCC dataset \citep{plasticc_data} is a simulation of transients observed by LSST under realistic observing conditions. The
full details of this simulation can be found in \citetalias{kessler19}. The PLAsTiCC dataset contains 3,500,734 light
curves \changeda{of} 18 different kinds of transient and variable sources. In contrast to the SNPhotCC \changeda{dataset,} which
only included different kinds of supernovae, the PLAsTiCC dataset also includes \changeda{other object types} such as variable stars,
micro-lensing events and active galactic nuclei. This introduces several challenges, as classifiers must be able
to handle more than just supernova-like objects. The details of all of the \changeda{object types} included in the
simulations are shown in Table~\ref{tab:plasticc_models} along with their counts.

\begin{deluxetable*}{rlrrrr}
\tablecaption{Summary of the \changeda{object types} included in the PLAsTiCC simulations
\citep{kessler19}. For each \changeda{object type}, the number of objects both with and without spectroscopic
confirmations is listed. The objects
with spectroscopic confirmations are a small fraction of the full sample, and they are not representative
of the distribution of the full sample.
\label{tab:plasticc_models}}
\tablehead{
    \colhead{ID\tablenotemark{a}} &
    \colhead{\changeda{Object type}} &
    \colhead{$N_{\textrm{confirmed}}$} &
    \colhead{$N_{\textrm{unconfirmed}}$} &
    \colhead{Galactic} &
    \colhead{Weight\tablenotemark{b}}
}
\startdata
90  & Type Ia SN                              & 2,313 & 1,659,831 & No & 1 \\
67  & Peculiar Type Ia SN -- 91bg-like        & 208   & 40,193    & No & 1 \\
52  & Peculiar Type Ia SN -- SNIax            & 183   & 63,664    & No & 1 \\        
42  & Type II SN                              & 1,193 & 1,000,150 & No & 1 \\
62  & Type Ibc SN                             & 484   & 175,094   & No & 1 \\
95  & Superluminous SN (Magnetar model)       & 175   & 35,782    & No & 1 \\
15  & Tidal disruption event                  & 495   & 13,555    & No & 2 \\
64  & Kilonova                                & 100   & 131       & No & 2 \\
88  & Active galactic nuclei                  & 370   & 101,424   & No & 1 \\
92  & RR Lyrae                                & 239   & 197,155   & Yes & 1 \\
65  & M-dwarf stellar flare                   & 981   & 93,494    & Yes & 1 \\
16  & Eclipsing binary stars                  & 924   & 96,472    & Yes & 1 \\
53  & Mira variables                          & 30    & 1,453     & Yes & 1 \\
6   & Microlens from single lens              & 151   & 1,303     & Yes & 1 \\
\hline
991\tablenotemark{c} & Microlens from binary lens              & 0     & 533       & Yes & 2 \\
992\tablenotemark{c} & Intermediate luminous optical transient & 0     & 1,702     & No & 2 \\
993\tablenotemark{c} & Calcium rich transient                  & 0     & 9,680     & No & 2 \\
994\tablenotemark{c} & Pair instability SN                     & 0     & 1,172     & No & 2 \\
\hline
    & Total                                   & 7,846 & 3,492,888 \\
\enddata
\tablenotetext{a}{Each \changeda{object type} was assigned a random ID number to identify it during the blinded phase of the
PLAsTiCC.}
\tablenotetext{b}{During the blinded phase of the PLAsTiCC, classifier performance was evaluated using
the metric defined Equation~\ref{eq:weighted_logloss_metric} with the class weights shown in this column.}
\tablenotetext{\changeda{c}}{These \changeda{object types} had no spectroscopically confirmed examples, and were included in the PLAsTiCC
to test anomaly detection algorithms. During the blinded phase of this challenge, they were all
assigned the same ID of 99 and treated as a single class.}
\end{deluxetable*}

Realistic observing conditions were simulated using the LSST Operations Simulator \citep{delgado14} for a three
year period of LSST operations. The \texttt{SNANA} package \citep{kessler09} was then used to simulate observations
for each of the included models following the generated observing conditions. A simulated trigger model is applied to
all of the generated transients following the DES supernova detection model \citep{kessler15}, and only objects
passing this trigger are kept.

Two distinct LSST survey components were simulated for the PLAsTiCC. The Wide-Fast-Deep (WFD) component
consists of observations covering almost half the sky. The Deep-Drilling-Fields (DDF) component consists of
5 different telescope pointings covering $\sim$50 deg$^2$. For the PLAsTiCC simulations, any observations
on the same night are co-added, so the DDF observations are effectively $\sim$1.5 mag deeper and $\sim$2.5
times more frequent than the WFD observations. There are significantly more observations in the WFD component,
and only 1\% of the objects passing the detection trigger are in the DDF sample.

The PLAsTiCC simulations include a model of the photometric and spectroscopic redshifts that will
be obtained for LSST. The simulations assume that \changeda{Galactic} objects can be cleanly separated from extragalactic
objects, and the measured photometric and spectroscopic redshifts of the \changeda{Galactic} objects are set to
zero. The simulation includes a model of a follow-up survey for extragalactic objects as described in
\citetalias{kessler19}. With this follow-up survey, 3.6\% of the extragalactic objects have spectroscopic
redshifts for their hosts. Extragalactic objects without spectroscopic
redshifts are assigned photometric redshifts and uncertainties on those photometric redshifts following a
model described in \citetalias{kessler19}.
In addition to spectroscopic redshifts, a total of 7,846 of the objects have spectroscopic confirmation
of their types, representing only 0.2\% of the total dataset. These
spectroscopically-classified objects are referred to as a ``training'' set for the rest of this article
as they are used to train the classifiers that will be applied to the ``test set'' of objects that
do not have spectroscopic classifications. 

\begin{figure}
    \epsscale{1.1}
    \plotone{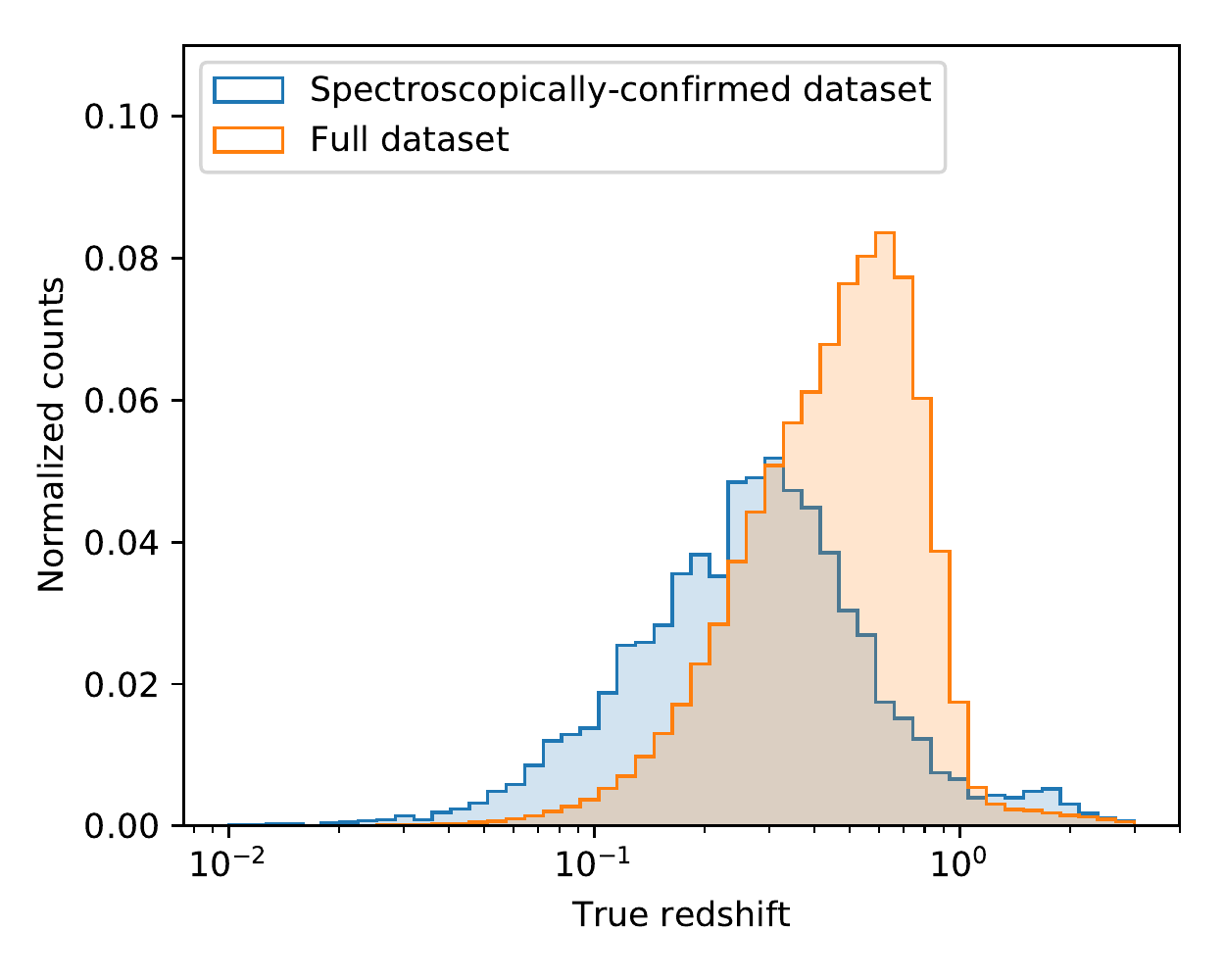}
    \caption{True redshift distributions for both the training and test sets in the PLAsTiCC dataset. The
    \changeda{simulated followup strategy for the} training dataset is strongly biased towards bright low-redshift objects.}
    \label{fig:redshift_bias}
\end{figure}

\begin{figure}
    \epsscale{1.1}
    \plotone{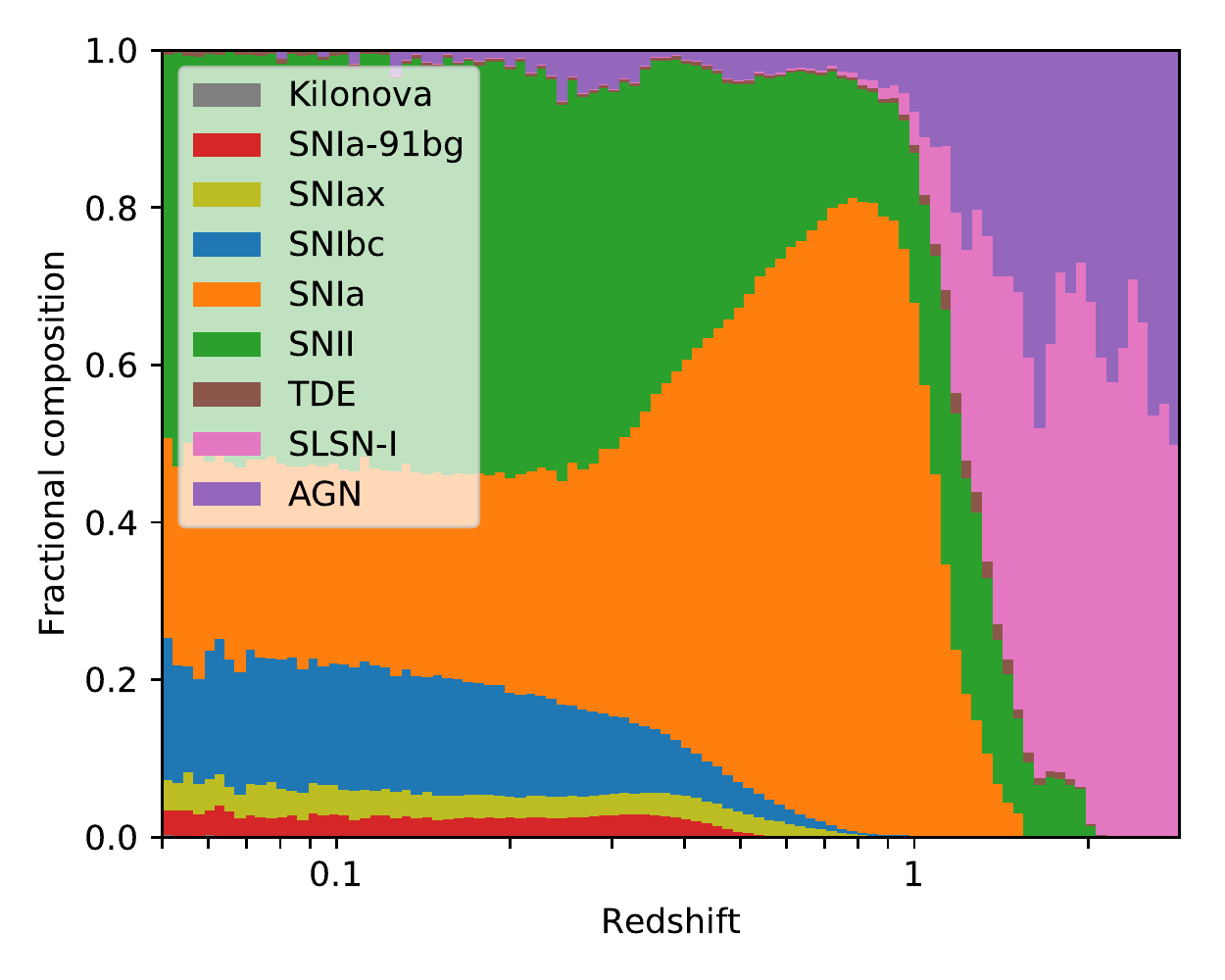}
    \caption{\changeda{Stacked histogram of the fraction} of objects belonging to each extragalactic object type as a
    function of redshift for the PLAsTiCC test dataset. The different object types have very different
    redshift distributions.}
    \label{fig:redshift_fractions}
\end{figure}

The training set for surveys such as LSST will \changeda{likely} be highly biased since spectra are required
to \changeda{categorize} each object in the training set. \changeda{Typical choices of followup strategies will}
preferentially select brighter, closer objects. For
the PLAsTiCC simulations, this bias can be seen in Figure~\ref{fig:redshift_bias}. The median redshift of
the training set is 0.18, compared to 0.43 for the full dataset. An additional challenge is that the different
transients and variables have very different redshift distributions, as illustrated in
Figure~\ref{fig:redshift_fractions}, so the biases in the training set will not be the same across different
object types.

\subsection{Metrics} \label{sec:metric}

\subsubsection{Flat-weighted metric}

As discussed in \citet{malz18}, the PLAsTiCC team proposed to evaluate the performance of different
classifiers on the PLAsTiCC dataset using a weighted multi-class logarithmic loss metric:

\begin{align}
    \label{eq:weighted_logloss_metric}
    \textrm{log. loss} =
    - \left(
    \frac{\sum_{i=1}^M w_i \sum_{j=1}^{N_i} \frac{y_{ij}}{N_i} \ln p_{ij}}{\sum_{i=1}^M w_i}
    \right)
\end{align}
where $M$ is the total number of classes and $N_i$ is the number of objects of each class.
$y_{ij}$ is 1 if the object $j$ belongs to class $i$ and 0 otherwise.
$p_{ij}$ are the predictions of a classifier, and for each object $j$ we should have
$\sum_j p_{ij} = 1$. The class weights $w_i$ can be chosen to emphasize the performance of the
classifier on specific classes.

An unweighted \changeda{logarithmic loss} is minimized when the classifier outputs predictions for each
class that match the conditional probabilities of each class given the observations. By dividing
by the total counts for each class, the \changeda{logarithmic loss} is normalized so that each class effectively
has the same weight. This is important for the PLAsTiCC simulations because some classes have
many more observations than others (e.g. over 10,000 SNe~Ia for each kilonova), and
an unweighted \changeda{logarithmic loss} would favor a classifier that does not attempt to classify the poorly-represented
classes. For an optimal classifier trained on this metric, $p_{ij}$ can be interpreted as a probability. By
this, we mean that given a \changeda{sample with} equal number of objects from each class, 40\% of the objects
that are assigned $p_{ij} = 0.4$ will belong to class $j$.

In our main analysis, we choose to use the metric of Equation~\ref{eq:weighted_logloss_metric} with class weights of
$w_i = 1$ for all of the classes present in the training set. We did not attempt to produce a classifier that
can identify objects that do not have examples in the training set, so we set $w_i = 0$ for all such classes
(the classes with IDs starting with 99 in Table~\ref{tab:plasticc_models}). We call this metric the
``flat-weighted metric'' because it gives all of the classes the same weights. The flat-weighted metric can be
written as:

\begin{equation}
    \label{eq:flat_weighted_metric}
    \textrm{Flat-weighted metric} = 
    - \left( 
    \frac{\sum_{i=1}^{T} \sum_{j=1}^{N_i} \frac{y_{ij}}{N_i} \ln p_{ij}}{T}
    \right)
\end{equation}
where the iteration $i$ over classes only considers classes that have examples in the training set,
and $T$ is the number of such classes.

\subsubsection{Redshift-weighted metric} \label{sec:redshift_weighted_metric}

For some science cases, a subtle issue with the flat-weighted metric in Equation~\ref{eq:flat_weighted_metric} is
that it does not take the redshift of extragalactic transients into account. If a classifier trained on this metric
is given information about the redshifts of the different objects, the classifier will learn to use the
redshift distributions of different transients in the training set to perform its classification. For example,
SNe~Ia tend to be discovered at higher redshifts than most other kinds of supernovae, so a classifier trained
on the flat-weighted metric will tend to classify ambiguous supernovae at high-redshifts as SNe~Ia, and ambiguous
supernovae at lower redshifts as other kinds of supernovae. Examples of this effect for the classifiers trained
in this paper will be shown in Section~\ref{sec:redshift_performance}.

For photometric classification, classifiers are typically trained on datasets of spectroscopically-confirmed
objects that have \changeda{biased} redshift distributions. It is the redshift distributions of these biased training sets
that will be encoded into the predictions, not the redshift distributions of the test set. For a classifier trained
on a metric similar to Equation~\ref{eq:weighted_logloss_metric}, any analysis that depends on understanding the
performance of the classifier as a function of redshift (such as cosmology with SNe~Ia) requires accurate estimates
of the differences between the redshift distributions in the training and full datasets. This is a \changeda{difficult
task if the spectroscopic followup is} distributed across many different telescopes with varying observing strategies and objectives.
One naive approach to dealing with this issue is to not input measurements of the redshifts of objects into the
classifier\changeda{, in an} attempt to prevent it from using the redshifts to make its decisions. However, the redshift affects
almost all features of a light curve, so a classifier can obtain a fairly accurate estimate of the redshift of an object
through features such as the relative flux levels in different bands or the time dilation of the light curve, and it
can use these estimated redshifts to make its predictions.

To mitigate this issue, we introduce the concept of a redshift-weighted metric. We reweight the objects in the
training set so that every object class effectively has the same chosen redshift distribution. When a classifier is
trained to optimize such a metric, the classifier cannot use the redshift distributions of the objects in the
training sample for classification because they are all identical. We implement such a metric by splitting the
redshift range into 10 logarithmically-spaced redshift bins between redshifts 0.1 and 3, along with an additional
bin for \changeda{Galactic} objects for a total of 11 redshift
bins. We then assign weights to each object to normalize the number of objects of each class in each redshift bin.
This results in the following redshift-weighted metric:

\begin{align}
    \label{eq:redshift_weighted_metric}
    \textrm{Redshift-}&\textrm{weighted metric} = \nonumber \\
    &- \left(
    \frac{\sum_{i=1}^M w_i \sum_{k=1}^{K} \sum_{j=1}^{N_i} \frac{y_{ijk}}{C_{ik}} \ln p_{ij}}
    {\sum_{i=1}^M w_i}
    \right)
\end{align}

Here, $y_{ijk}$ is 1 if object $i$ belongs to class $j$ and is in redshift bin $k$. $K$ is the total number of redshift bins.
$C_{ik}$ is the total number of objects in class $i$ and redshift bin $k$. To avoid extremely large weights for objects in
bins that have very few counts, we impose a floor on $C_{ik}$ of 100 objects. For extragalactic objects, we choose
to set the weight to $w_i = 1$. A typical extragalactic class is well-represented in roughly half of the
different redshift bins while \changeda{Galactic} objects only have a single bin. To roughly maintain a flat class weighting,
we set the weight for \changeda{Galactic} objects to the average number of bins that are well-populated for extragalactic classes
($\sim$5). As for the flat-weighted metric, we set $w_i = 0$ for the objects that were not present in the training set
since we did not attempt to classify them.

\subsubsection{Kaggle metric}

For the blinded phase of the PLAsTiCC hosted on Kaggle, the performance of classifiers was evaluated using the ``Kaggle
metric'' which is of the weighted multiclass logarithmic-loss metric in Equation~\ref{eq:weighted_logloss_metric}
with the class weights shown in Table~\ref{tab:plasticc_models}. One part of the blinded phase of the PLAsTiCC was
identifying new kinds of objects that had no examples in the training set (the ``class 99'' objects). In this work,
we did not attempt to address this part of the challenge. \changeda{Use of the Kaggle metric requires values for the class 99
objects. Hence, }to evaluate the performance of our classifiers on the Kaggle metric, we generate \changeda{artificial}
predictions for the class 99 objects using \changeda{formulae} that
were tuned \changeda{by probing} the metric during the blinded phase of the PLAsTiCC. For \changeda{Galactic} objects, we
assign a flat predicted probability of 4\% to the class 99 objects. For \changeda{each extragalactic object}, we assign a predicted
probability according to the following formula:

\begin{equation}
    \label{eq:class_99_prediction}
    P_{99} = P_{42} + 0.6 \cdot P_{62} + 0.2 \cdot P_{52} + 0.2 \cdot P_{95}
\end{equation}

where $P_X$ is the predicted probability assigned to the class with ID $X$ (see Table~\ref{tab:plasticc_models} for the
list of IDs). We then rescale all of the predicted probabilities so that they sum to 1 for each object. Note that these
\changeda{formulae} are not a proper way of identifying new objects in the data. All of the top 5 performing teams in the blinded
phase of the PLAsTiCC used similar \changeda{formulae}, and we are not aware of any successful attempts \changeda{to identify} new
objects. \changeda{For the analyses in this paper,} we primarily use the flat-weighted metric and redshift-weighted metric\changeda{, which} both
ignore the class 99 objects. We do however evaluate our performance on the Kaggle metric using these \changeda{formulae} for the class
99 predictions for comparison purposes.

\subsubsection{Single class metrics} \label{sec:single_class_metrics}

Finally, we evaluate several standard metrics for the performance of \changeda{a deterministic classifier when identifying
objects of a single specific type.} The confusion matrix and corresponding labels
for each of the outcomes of classification of one transient type out of a larger sample are shown in Table
\ref{tab:binary_classification}. Using the labels from this table, we define the following metrics that will be used
in further analysis:

\begin{equation}
    \textrm{true positive rate (TPR)} = \frac{\textrm{TP}}{\textrm{TP} + \textrm{FN}}
\end{equation}

\begin{equation}
    \textrm{false positive rate (FPR)} = \frac{\textrm{FP}}{\textrm{FP} + \textrm{TN}}
\end{equation}

\begin{equation}
    \textrm{purity} = \frac{\textrm{\changeda{TP}}}{\textrm{TP} + \textrm{FP}}
\end{equation}

\begin{equation}
    \textrm{completeness} = \frac{\textrm{TP}}{\textrm{TP} + \textrm{FN}}
\end{equation}

\begin{deluxetable}{lc|cc}
\tablecaption{Confusion matrix for classification of a single \changedb{object type} (P) out of a larger sample of other \changedb{object types} (N).
\label{tab:binary_classification}}
\tablehead{
    & & \multicolumn{2}{c}{True Class} \\ 
    & & \colhead{P} & \colhead{N}
}
\startdata
Predicted & P & True positive (TP) & False positive (FP) \\ 
Class & N & False negative (FN) & True negative (TN) \\
\enddata
\end{deluxetable}

We also calculate the Area under the Receiver Operator Characteristic Curve (AUC) for each of our classes. This
metric is defined as the area under the curve of the TPR plotted against the FPR, and ranges between 0.5 for a
random classifier to 1 for a perfect classifier. See \citetalias{lochner16} for more complete definitions of
all of these metrics.

\section{Methods} \label{sec:methods}

\subsection{Overview} \label{sec:overview}

Our approach to photometric classification combines several techniques. \changeda{We first preprocess the light curves as
described in section~\ref{sec:preprocessing}.} In Section~\ref{sec:gaussianprocessfit},
we describe how we use Gaussian process (GP) regression to predict smooth models for each of our sparsely
sampled light curves. Using these GP models, as discussed in Section~\ref{sec:augmenting}, we augment the
spectroscopically-confirmed dataset, generating \changeda{artificial} light curves that are more representative of the
full dataset. In Section~\ref{sec:features}, we describe how for each light curve in the augmented training
set, we calculate a set of features from the GP models. We then train a tree-based classifier on the extracted
features to perform the final classification predictions, the details of which can be found in
Section~\ref{sec:lightgbm}.

\subsection{Light curve preprocessing} \label{sec:preprocessing}

The fluxes of transients are typically determined by subtracting newly measured fluxes from fluxes measured
from a set of reference images. For long-lived transients, these reference images may contain light from
the transients themselves. \changeda{The blinded PLAsTiCC dataset} did not provide the reference fluxes
of the sources,
so for objects such as variable stars, the ``background'' level of the light curve is simply the flux
of the light curve at an arbitrary point in time. To address this issue, we estimate new ``background''
levels for each light curve using a biweight estimator \citep{beers90}. For short-lived
transients, this background estimator will return the flux level at times when there is no light from
the transient. For sources such as variable stars or active galactic nuclei, this robust estimator will
effectively return the mean value of the light curve.

\subsection{Modeling light curves with Gaussian process regression} \label{sec:gaussianprocessfit}

Gaussian process (GP) regression has been shown to be effective for several applications of astronomical
light curve modeling. \citet{kim13} used GPs to model the light curves of SNe~Ia and predict their peak
brightnesses. \citet{fakhouri15} \changeda{and \citet{saunders18}} modeled the full spectral time-series of SNe~Ia with GPs, and used these
models to evaluate the spectra of these objects at arbitrary times. \citetalias{lochner16} introduced
GP modeling for astronomical transient classification. \citetalias{revsbech18} showed that GP models can be used to
augment a biased training set by generating additional training data from the GPs. These works all
focused specifically on using GPs to model particular kinds of supernovae. We extend these \changeda{techniques} so that
they can be applied to a wider range of transients and variables. An introduction to GP regression can be found
in Appendix~\ref{appendix:gaussianprocess}.

Previous works using GPs for photometric classification (e.g. \citetalias{lochner16} and \citetalias{revsbech18})
evaluated \changeda{separate GPs} for each band of the light curve, so the model was not able to take cross-band
information into account. In contrast, in this work, we use GPs in both time and wavelength to model the
light curve in all bands simultaneously. We do not attempt to explicitly model the throughputs of the
different filters. Instead, we calculate central wavelengths for each of the bands using the estimated LSST
throughputs\footnote{\url{https://github.com/lsst/throughputs}}
assuming a source with a constant $F_\lambda$ spectrum. We use these central wavelengths as the
coordinates for the wavelength dimension of the GP. This effectively means that the GP is producing a model of
the spectrum convolved with a broad filter rather than modeling the spectrum directly.

We use the \texttt{George} package \citep{ambikasaran15} to implement our GPs.
Using maximum likelihood estimation, we fit for both the amplitude ($\alpha$) and time length scale ($l_t$)
parameters on a per-object basis. It is difficult to reliably fit the length scale in wavelength on a
per-object basis due to the fact that there are only 6 filters used for observations. We choose to fix the
length scale in wavelength to 6000\AA~because we find that this value produces reasonable models for all of
the transients and variables in the PLAsTiCC dataset. \changeda{The results of this analysis are not highly
sensitive to the choice of the length scale in wavelength.}

\begin{figure*}
    \epsscale{1.15}
    \plottwo{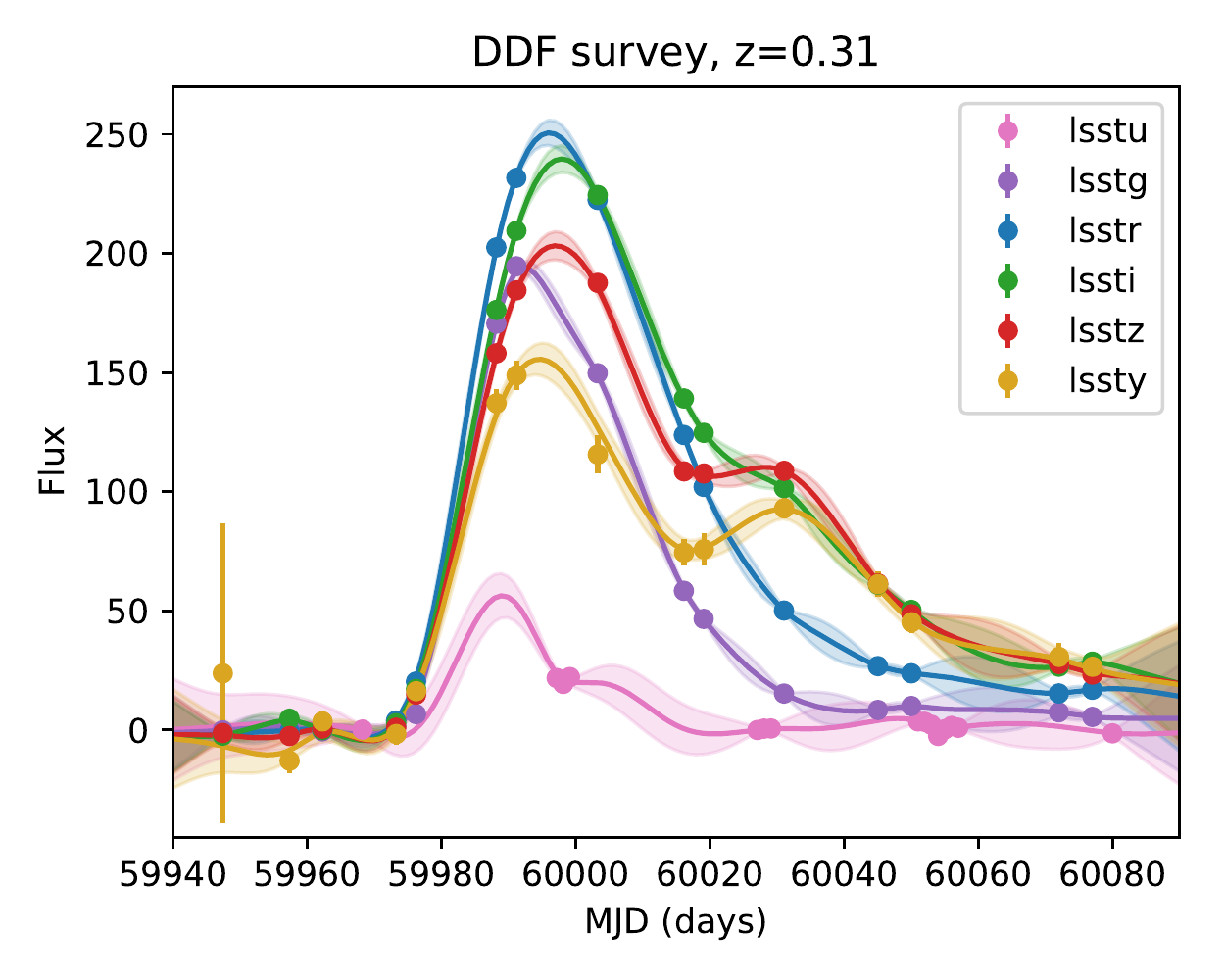}{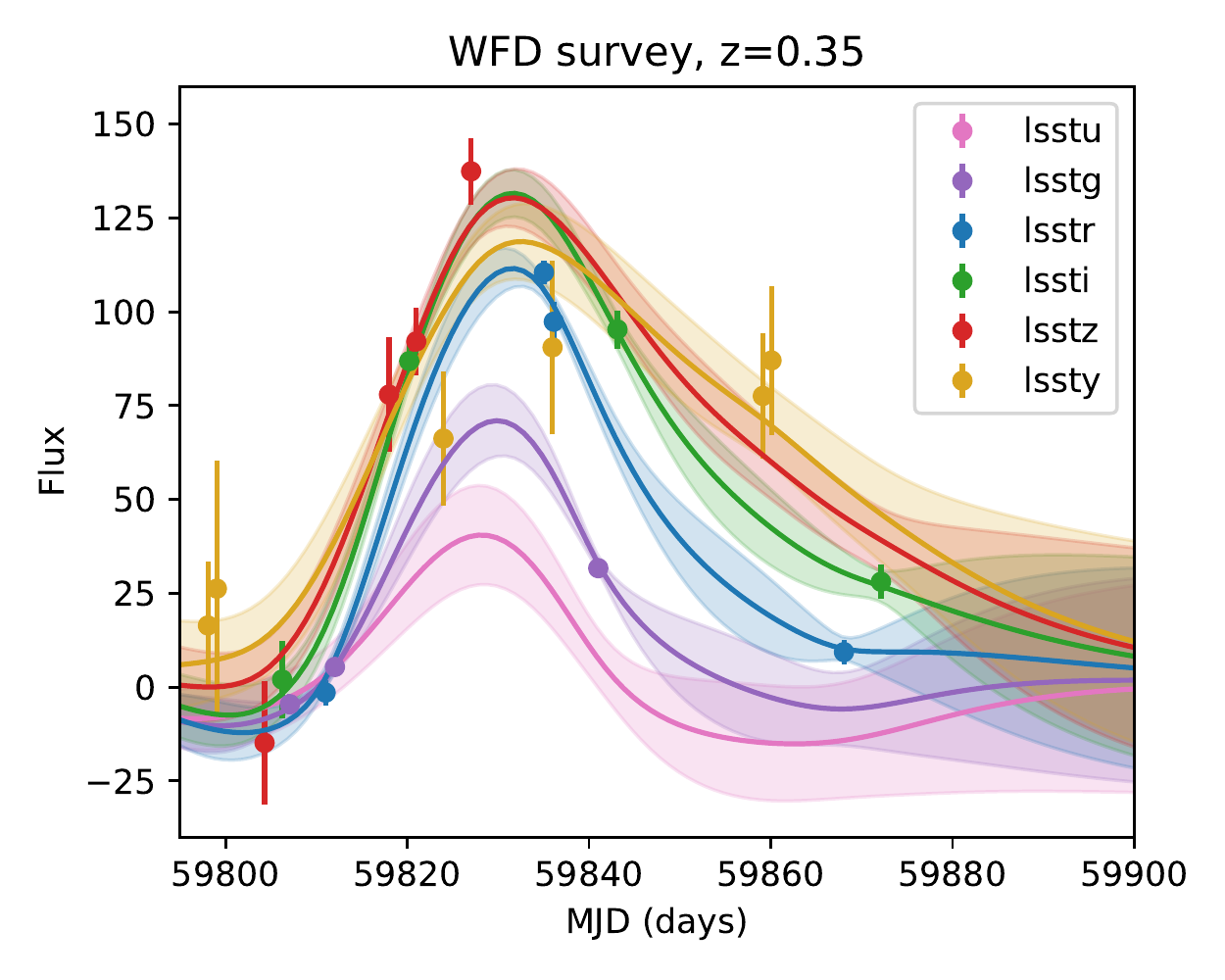}
    \caption{Examples of GP models for SN~Ia light curves. Left: A well-sampled,
    high signal-to-noise light curve. Right: a poorly-sampled, lower signal-to-noise light curve. The
    mean GP flux prediction for each band is shown as a solid line surrounded by a shaded contour indicating the
    one-standard-deviation uncertainty on the flux prediction.}
    \label{fig:gp_snia}
\end{figure*}

Examples of the GP model for a well-sampled SN~Ia light curve and a poorly sampled one are shown in
Figure~\ref{fig:gp_snia}. The GP produces reasonable non-parametric models of the light curves that can be used for further
analysis, along with estimates of the model uncertainties. Because we are using a kernel in both wavelength and time,
the GP is able to use cross-band information to infer the supernova light curve even in bands where there are
few observations. This can be seen in the right plot of Figure~\ref{fig:gp_snia} where the GP produces a reasonable
model (with high uncertainty) of the light curve in the LSST~$u$ band even though there are no observations in this
band. 

\subsection{Augmenting the training dataset} \label{sec:augmenting}

\changedc{The spectroscopically-classified objects that are used as a training set for photometric
classifiers} tend to be highly non-representative of the full dataset \changeda{in terms of their redshift
and signal-to-noise distributions}. These training sets are typically strongly
biased towards bright, low-redshift objects. Most previous attempts at producing photometric classifiers have
seen strongly degraded performance when trained on non-representative datasets, and have concluded that
obtaining representative training sets is essential for photometric classification (e.g. \citetalias{lochner16}).
More recently, \citetalias{revsbech18} showed that it is possible to apply various transformations to the light
curves in the original training set to generate a new training set that is more representative of the test set.
We call this process ``augmentation'' of the training set. Using their \texttt{STACCATO} framework for augmentation,
\citetalias{revsbech18} train a classifier whose performance is significantly better (AUC of 0.96 \changeda{on the
SNPhotCC dataset}) than one trained on the original non-representative training set (AUC of 0.93), and
approaching the performance of a classifier trained on a representative training set (AUC of 0.977).

In \texttt{STACCATO}, GPs are fit to the observations of each object in the training set, with separate GPs for each band.
``Synthetic light curves'' are then produced for each object by sampling from the GPs. Each sample from a GP produces a different
continuous function that can be interpreted as a synthetic light curve that is consistent with the observations of the
original object. By repeatedly sampling from the GPs, many synthetic light curves can be produced for each object
in the training set. In \texttt{STACCATO}, for every object, a ``propensity score'' is calculated, which is an estimate
of how likely the object is to make it into the training set. The propensity score is then used to determine how many different
synthetic light curves to make for each object in the training set. By generating different number of synthetic light curves
for each object in the training set, an ``augmented'' training set of synthetic light curves is produced that is more
representative of the test set. Finally, a classifier is trained on the set of synthetic light curves.

Our approach to augmentation differs in several ways from the approach of \citetalias{revsbech18}. In \texttt{STACCATO}, different
synthetic light curves are generated for each object, but these synthetic light curves all use the same set of observations.
Instead, our augmentation procedure involves simulating entirely new sets of observations for each object. \changeda{
When augmenting a light curve from the training set, we throw out large blocks of observations to simulate season boundaries,
take originally well-sampled light curves and degrade their sampling, and add noise to the light curve in different bands. 
We measure the cadence and depth of observations in the test set, and generate augmented light curves that have similar
cadences and depths of observations to the test set. This ensures that the light curves in the augmented training set have
observations that are representative of the observations of light curves in the test set regardless of the light curve quality
in the original training set. We also interpret the GP uncertainty as an uncertainty due to poor measurement rather than
intrinsic variation of the light curve. For this reason, we choose to use the mean prediction of the GP for our augmented
light curves, and we propagate the GP prediction uncertainties into the uncertainties of the generated observations.
}

Additionally, we introduce the concept of ``redshift augmentation'' where we take an object in the training set and simulate
observations of it at different redshifts. Because we are using a Gaussian process in both time and wavelength, we can
shift the redshift of an object by \changeda{evaluating the Gaussian process predictions for the light curve} of that object at the
redshifted wavelengths. Note that the GP is effectively modeling the spectrum of the object convolved with a filter.
Assuming that this convolved spectrum is reasonably smooth and that the different filters have similar profiles, \changeda{
the two-dimensional GP predictions will automatically include} k-corrections \citep{oke68} to the observed brightnesses in each
filter. There will be higher-order corrections due to sharp structure in the spectrum (e.g. emission lines)
and differences in the filter shape, but these are unlikely to significantly affect the classification in most cases. When
redshifting the observations of an object, we then update the observed brightnesses by calculating the difference
in distance modulus assuming a fiducial cosmology. After these procedures, we have effectively simulated the light curve for
an object at a different redshift than it was originally observed at.

With redshift augmentation, if we observe an object at one redshift, then we can effectively use that object for
training at all redshifts. Because training samples are typically biased towards bright, low-redshift objects,
this means that we can use redshift augmentation to fill in the missing regions of parameter space in the training
sample at high redshifts. This differs from the augmentation procedure in \texttt{STACCATO}. While \texttt{STACCATO} is making additional
versions of light curves for objects that were already at high redshifts, we are \changeda{instead taking low-redshift}
light curves and simulating what they would look like if they had been observed at high redshifts. \changeda{A potential
caveat with redshift augmentation is that the subpopulations of different object types could evolve with redshift.
We discuss how this this can be addressed in Section~\ref{sec:redshift_drift}.}

One major challenge with augmentation is determining where in parameter space to generate new objects to match the
training set to the test set. \texttt{STACCATO} uses a ``propensity score'' to decide how many new versions should be generated
for each object in the training set. Unfortunately, for this \changeda{technique} to be effective, the rates and selection efficiencies
in \changeda{the} test set must be known for each object type. The rates of different transients are not currently well known in many
regions of parameter space. For SNe~Ia and core-collapse supernovae, the current best measurements of the rates above
redshift 1 have uncertainties of roughly 50\% of the measured rates (e.g. \citet{okumura14, rodney14, strolger15}). To address
this issue, we instead choose to design a classifier whose performance is independent of the rates and selection efficiencies
in the training set. This can be done by training a classifier to optimize the metric described in
Section~\ref{sec:redshift_weighted_metric}\changeda{; a} discussion of the effectiveness of this procedure will be shown in
Section~\ref{sec:redshift_performance}. \changeda{When training} a classifier with these properties, for the augmentation procedure,
we simply need to ensure that we generate a set of light curves covering the full parameter space for any object type at a given
redshift. We therefore simulate each object in our training set the same number of times at a range of different redshifts.

The full details of our augmentation procedure can be found in
Appendix~\ref{sec:appendix_augment}. A summary of our approach to augmentation is as follows:
\begin{enumerate}
    \item Fit a GP to a light curve in the original training sample to use as a template.
    \item Choose a new redshift for extragalactic objects, or brightness for \changeda{Galactic} objects.
    \item \changeda{Evaluate the mean GP prediction and uncertainty} to obtain a new light curve at the chosen brightness/redshift.
    \item Drop observations to \changeda{simulate} poorly-sampled light curves from the well-sampled training
        light curves.
    \item Add measurement uncertainties that are representative of the test dataset.
    \item Ensure that the generated light curve would be detected.
    \item Simulate a photometric redshift following the mapping between spectroscopic
        and photometric redshifts in the test dataset.
    \item Repeat these steps until a large enough augmented sample has been obtained.
\end{enumerate}

For this analysis, we use the same augmented training set to train all of the different classifiers that will
be discussed. We generate up to 100 different versions of each light curve in the training set\changeda{,} which results in
a total of 591,410 light curves in the augmented training set.

\subsection{Extracting features from the light curves} \label{sec:features}

To extract features from a light curve, we begin by performing a GP fit as described in
Section~\ref{sec:gaussianprocessfit} and computing the mean GP flux predictions in the LSST~$g$, $i$,
and $y$ bands. We choose to use the observer-frame LSST~$i$ band as the reference for many of our
features because this band typically has a reasonable flux level for both the low and high-redshift
objects in our sample. We extract a variety of features from the GP flux predictions in each of these
three bands, the details of which can be found in Table~\ref{tab:features}.

\startlongtable
\begin{deluxetable*}{p{1.5in\hangindent=1em}p{5in}}
\tablecaption{Overview of features used for classification. Unless specified otherwise, all features
are calculated using the mean flux predictions of a GP fit to the light curve. \label{tab:features}}
\tablehead{
    \colhead{Feature name} &
    \colhead{Description}
}
\startdata
    \texttt{host\_photoz} & Host-galaxy photometric redshift, taken directly from the PLAsTiCC metadata. \\
    \texttt{host\_photoz\_err} & Host-galaxy photometric redshift error, taken directly from the PLAsTiCC metadata. \\
    \texttt{length\_scale} & Fitted GP length scale hyperparameter, in days. \\
    \texttt{max\_mag} & Peak magnitude of the GP flux prediction in the LSST $i$ band. \\
    \texttt{pos\_flux\_ratio} & Ratio of the maximum positive flux to the maximum-minus-minimum flux in the LSST $i$ band. \\
    \texttt{[max,min]\_flux\_ratio \_[blue,red]} & Normalized difference of the light curve colors at maximum/minimum light. The blue
        measurement is the difference between the LSST $i$ and $g$ bands, and the red measurement is the difference between the
        LSST $y$ and $i$ bands. The normalized difference is calculated by taking the difference of the fluxes in the two bands
        divided by their sum. \\
    \texttt{max\_dt} & Difference of the time of maximum in the LSST $y$ and $g$ bands in days. \\
    \texttt{[positive,negative] \_width} & An estimate of the light curve ``width'' that is applicable even for non-supernova-like
        transients and variables. This is implemented as the integral of the positive/negative parts of the GP
        flux predictions divided by the positive/negative maximum fluxes. \\
    \texttt{time\_[fwd,bwd]\_max \_[0.2,0.5]} & Measurements of the rise and decline times of a light curve. This measurement is
        defined as the time in days for the light curve to rise (bwd) or decline (fwd) to a given fraction
        (either 20\% or 50\%) of maximum light in the LSST $i$ band. \\
    \texttt{time\_[fwd,bwd]\_max \_[0.2,0.5]\_ratio \_[blue,red]} & Ratio of the rise/decline times calculated as described above
        in different bands. The blue measurement is the difference between the LSST $i$ and $g$ bands, and the red measurement
        is the difference between the LSST $y$ and $i$ bands. \\
    \texttt{frac\_s2n\_[5,-5]} & Fraction of observations that have a \changeda{signal greater than 5/less
        than -5 times the noise level.} \\
    \texttt{frac\_background} & Fraction of observations that have an absolute signal-to-noise less than 3. \\
    \texttt{time\_width\_s2n\_5} & Time difference in days between the first observation with a signal-to-noise greater than 5
        and the last such observation (in any band). \\
    \texttt{count\_max\_center} & Number of observations in any band within 5 days of maximum light. \\
    \texttt{count\_max\_rise \_[20,50,100]} & Number of observations in any band between 20, 50, or 100 days before maximum
        light and 5 days after maximum light. \\
    \texttt{count\_max\_fall \_[20,50,100]} & Number of observations in any band between 5 days before maximum light and
        20, 50, or 100 days after maximum light. \\
    \texttt{peak\_frac\_2} & Ratio of the maximum flux in the second most prominent peak in the light curve to the maximum
        flux in the main peak, averaged over all LSST bands. This is intended to identify supernova-like objects that only
        have a single large peak in most bands. \\
    \texttt{total\_s2n} & Total signal-to-noise of all observations of the object. \\
    \texttt{percentile\_diff \_[10,30,70,90]\_50} & Measurements of the distributions of the observed fluxes. For each band,
        we calculate the flux level for a given percentile of observations, and normalize it by the maximum-minus-minimum of the
        GP flux predictions. We then take differences between this measurement for various percentiles and
        the $50^{th}$ percentile measurement. The final value is the median of the calculated differences in all bands. \\
\enddata
\end{deluxetable*}

In general, we find that in the PLAsTiCC dataset, the non-supernova-like variables and transients end
up being relatively easy to distinguish, so most our our effort went to identifying features that are
effective for distinguishing the different kinds of supernovae. We initially generated hundreds of different
features, and we used both the feature importance ranking of our classifier and cross-validation
performance (discussed in Section~\ref{sec:lightgbm}) to select a subset of 41 features that give
good performance. Most of the features that we include are standard features that are well-known to distinguish transients,
such as the \changeda{apparent} peak brightness of the transient and the photometric redshift. We include measures of
the colors of the objects by taking ratios of the peak brightnesses in different bands, and estimates of
the rise time and fall times that help distinguish between the different supernova types.

One major difference between this analysis and most previous ones is that our light curve model includes
cross-band information. \changedb{The GP flux predictions in a given band incorporate information from nearby bands
because of the GP kernel in the wavelength direction. We only calculate features from the GP predictions in the
LSST-$g$, $i$, and $y$ bands, but these features capture the observations in all of the different bands.}
This can be seen for the poorly-sampled light curve in Figure~\ref{fig:gp_snia} where we have reasonable models of the
light curve even in bands with no observations. We find that calculating features off of GP flux predictions in
additional bands beyond the three previously listed does not improve the classification for the PLAsTiCC sample.
This kind of analysis is only possible with a light curve model that fits both wavelength and time directions
simultaneously as opposed to traditional models where each band is fit independently. Although we chose to
perform the GP flux predictions at the wavelengths associated with several LSST bands for simplicity, the
GP flux predictions can be performed at arbitrary wavelengths, and observations from different bands or other
telescopes can easily be included in the GP model.

There are a handful of notable features that we introduced that have not been included in previous analyses.
First, we add several features that are effective at classifying variable object types, most notably the
\texttt{percentile\_diff\_X\_Y} features that measure the distribution of the photometry values. These
features are effective at distinguishing object types with large wings in their photometry distributions
(e.g. eclipsing binaries) from object types that have more even distributions (e.g. Mira variables),
even with relatively few observations.

\begin{figure*}
    \epsscale{1.15}
    \plottwo{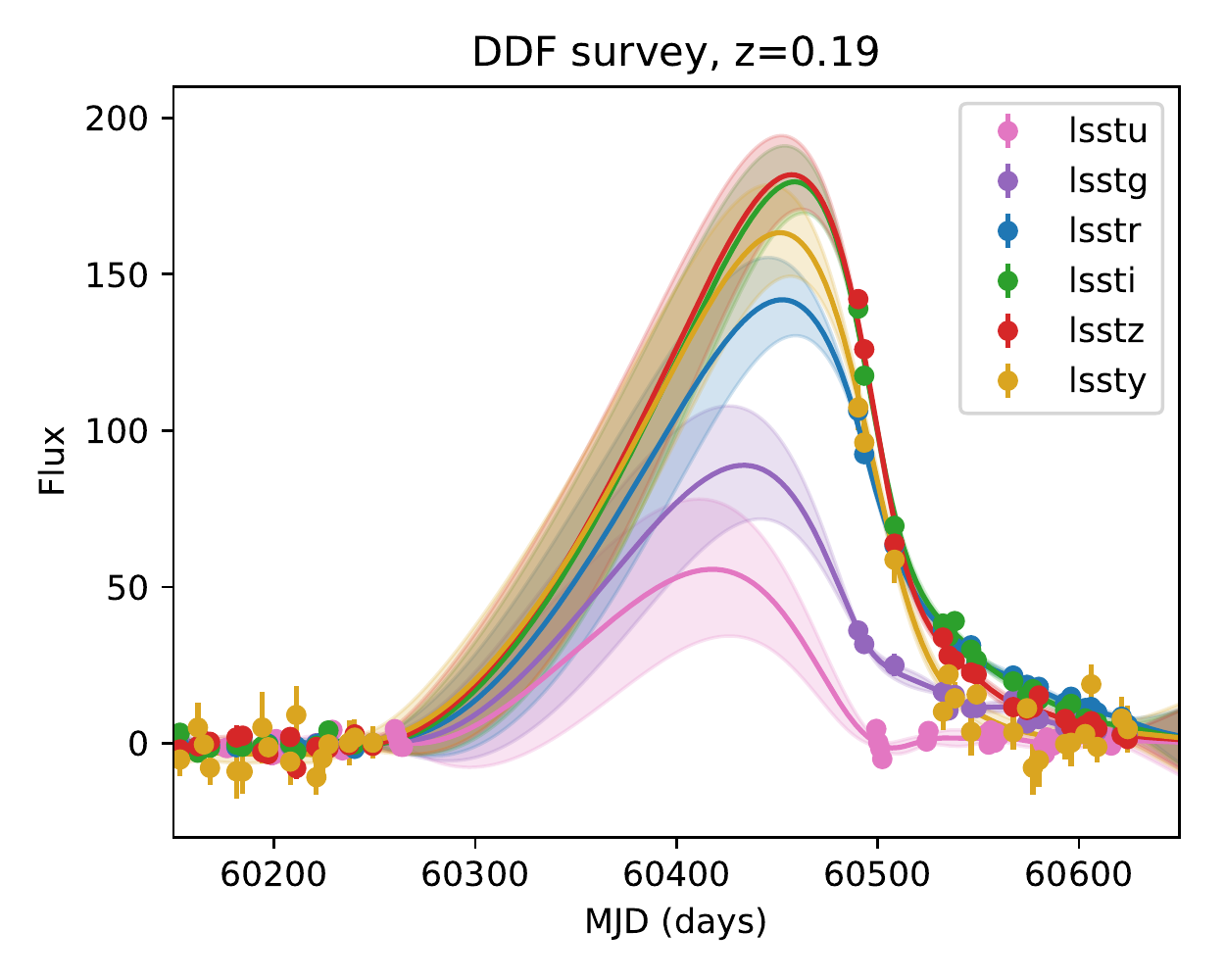}{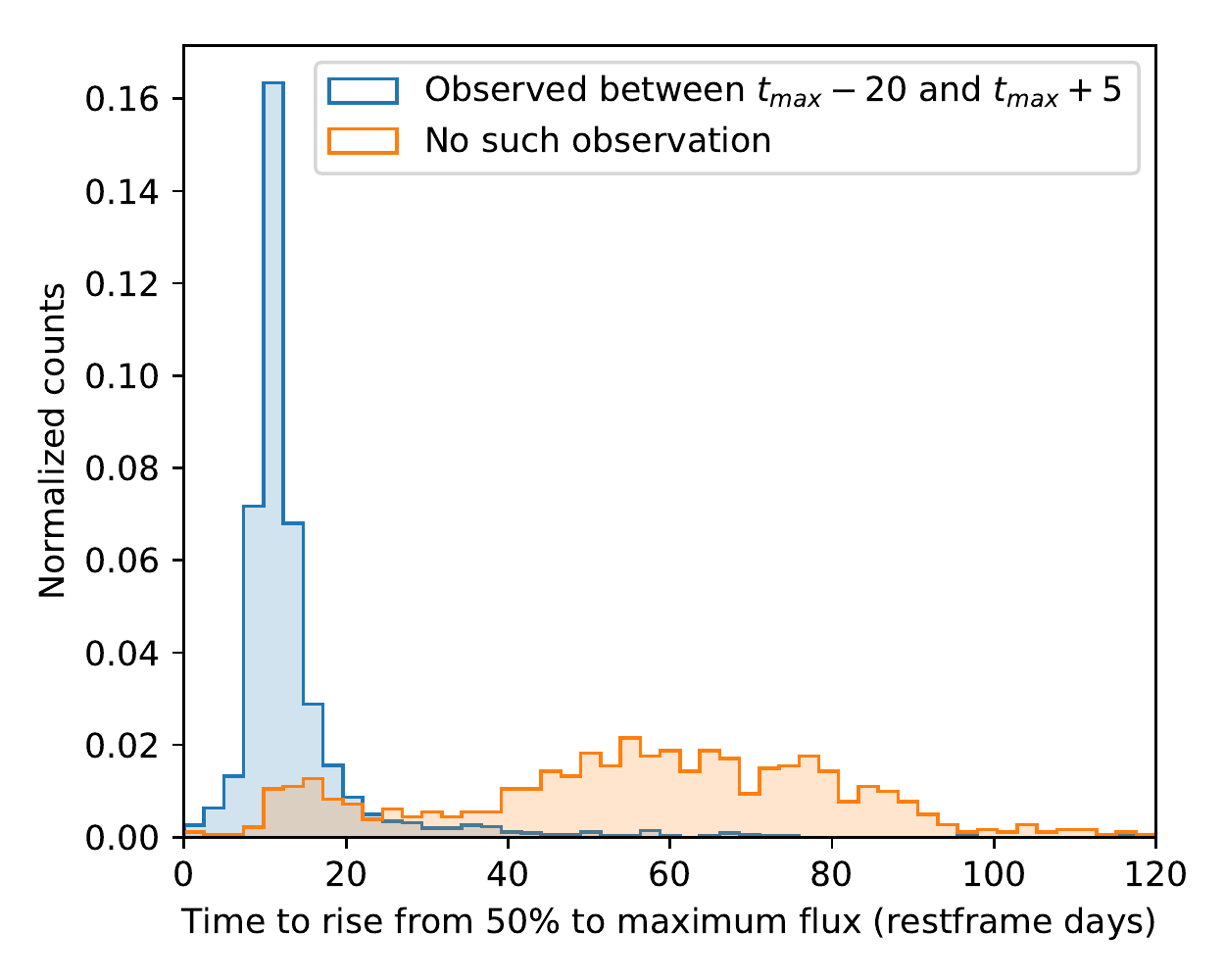}
    \caption{\changedb{Example of a feature that probes the fit quality. Left panel:} Example of the GP model
    for a SN~Ia light curve with
    no data before maximum light. See Figure~\ref{fig:gp_snia} for details of the plot. The GP
    is unable to constrain the rise time, and produces a model with large uncertainties.
    \changedb{Right panel}: Histogram of the measured rise times for SNe~Ia. The rise time is well-constrained
    (blue histogram) for objects with an observation between 20 observer-frame days before maximum light
    and 5 days after maximum light. Objects without such an observation (orange histogram) have rise time
    measurements that are biased to higher values with large dispersions.}
    \label{fig:risetime_counts}
\end{figure*}

Another novel \changeda{technique} in this work is the introduction of features that measure the \changeda{fit quality}. In
many cases, the available photometry for an object does not cover its full light curve. An example of
such a light curve for a Type~Ia supernova can be seen in Figure~\ref{fig:risetime_counts}. Without
any photometry before maximum light, the GP produces a model with large uncertainties,
and the measurement of the rise time will be both uncertain and biased. We experimented with several
approaches to incorporating this information into the classifier, and found that measuring the number
of observations in various bands around maximum light provides an excellent way to evaluate the
accuracy of the rise time measurements. The right panel of Figure~\ref{fig:risetime_counts} illustrates
this for SNe~Ia: light curves that have an observation between 20 observer-frame days before
maximum light and 5 days after maximum light have a tight distribution of measured rise times, while
light curves without this information have a wide, biased distribution of rise times. By adding features measuring
the number of observations in various time intervals, the classifier can effectively determine how
reliable other features, such as the rise time information, are.

\subsection{Training a LightGBM model} \label{sec:lightgbm}

We train a gradient boosted decision tree classifier on features extracted from the augmented training set.
Decision trees are a classification technique where objects are filtered through a variety of cuts to attempt
to separate them into their different classes. Boosted decision trees combine a large number of these trees
to produce a robust classifier, and have proven to be very effective at a variety of different classification
tasks in astronomy (e.g: \citet{bailey07}, \citetalias{lochner16}). Using the LightGBM implementation of
gradient boosted decision trees \citep{ke17}, we train separate classifiers to optimize
objective functions that are direct implementations of the metrics in Equations~\ref{eq:weighted_logloss_metric}
and \ref{eq:redshift_weighted_metric}.

To evaluate and optimize the performance of the classifier, we use five-fold cross-validation on the augmented
training set. We partition the augmented training data into five separate subsets that each have equal ratios of the
different targets. We then train five separate classifiers, each of which is trained on four of the five
subsets, and we evaluate its performance on the remaining subset. By repeating this procedure for each of the
subsets, we obtain out-of-sample predictions for every object in the augmented training set. We then evaluate the
PLAsTiCC metric on these predictions, and use this cross-validation performance to tune our model. For the
augmented dataset, we ensure that all light curves generated from the same original light curve are included
in the same subset to avoid leaking information across the subset boundaries. When generating predictions for
objects that are in the test set, we evaluate the average of the classification probabilities for the new
objects from each of the five trained classifiers.

We optimize the \changeda{hyperparameter values} of the LightGBM model by scanning over each \changeda{hyperparameter} individually and evaluating
the cross-validation performance \changeda{on the flat-weighted metric}. The resulting \changeda{hyperparameter} values are shown in
Table~\ref{tab:lightgbm_parameters}. The optimal \changeda{hyperparameter values} are relatively stable across different sets of
features and target metrics, so for simplicity we use the same \changeda{hyperparameter values} for all of the analysis variants.

\begin{deluxetable}{ll}
\tablecaption{Optimized \changeda{hyperparameter values} used for the LightGBM model.
\label{tab:lightgbm_parameters}}
\tablehead{
    \colhead{\changeda{Hyperparameter} name} &
    \colhead{Value}
}
\startdata
    \texttt{boosting\_type} & \texttt{gbdt} \\
    \texttt{learning\_rate} & \texttt{multi\_logloss} \\
    \texttt{colsample\_bytree} & 0.05 \\
    \texttt{reg\_alpha} & 0 \\
    \texttt{reg\_lambda} & 0 \\
    \texttt{min\_split\_gain} & 10 \\
    \texttt{min\_child\_weight} & 2000 \\
    \texttt{max\_depth} & 7 \\
    \texttt{num\_leaves} & 50 \\
    \texttt{early\_stopping\_rounds} & 50 \\
\enddata
\end{deluxetable}

We train two separate versions of our classifier, one of which is optimized for performance on the
flat-weighted metric defined in Equation~\ref{eq:flat_weighted_metric}, and one of which is optimized
for performance on the redshift-weighted metric defined in Equation~\ref{eq:redshift_weighted_metric}.
Both of these classifiers are trained on the same augmented training set. \changeda{For the training set,
the Kaggle metric is nearly identical to the flat-weighted metric, so we do not train a separate
classifier to optimize it.}

\begin{figure*}
    \epsscale{1.15}
    \plotone{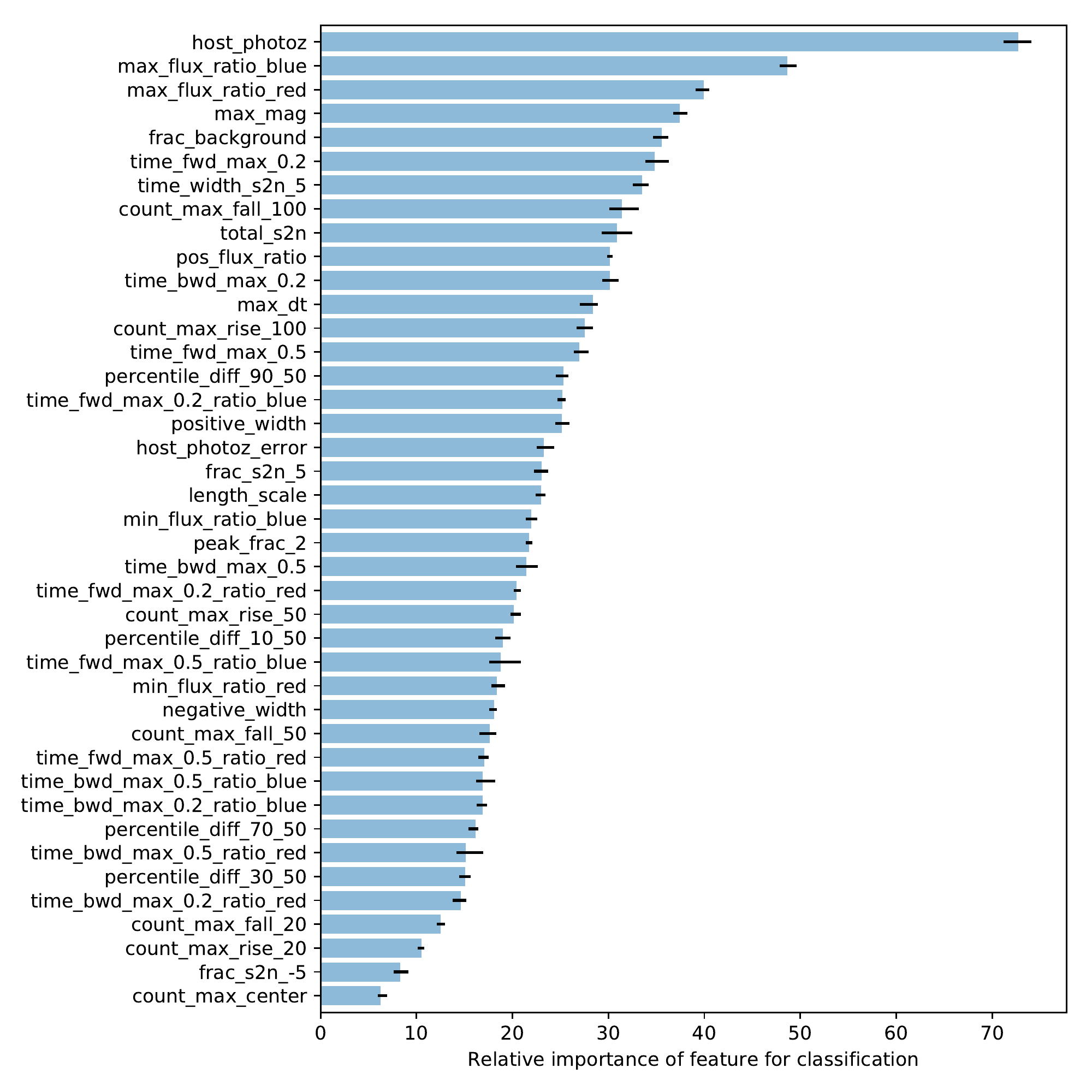}
    \caption{Relative importance of each feature for classification for a classifier trained to optimize the
    flat-weighted metric. Definitions of these features can be found in Table~\ref{tab:features}. Five separate
    classifiers were trained on different folds of the training set. The blue bars represent the mean importance of
    each feature across the five classifiers. The thin black bars indicate the range of importance across the
    five classifiers.}
    \label{fig:feature_importances}
\end{figure*}

LightGBM outputs a measure of how much each feature contributed to the classification. We call this measure the
``importance'' of that feature for classification. The importance of each feature for a classifier trained
to optimize the flat-weighted metric are shown in Figure~\ref{fig:feature_importances}. The most important
features for classification are the photometric redshift of the host galaxy, the peak brightness, and the colors
of the light curves at maximum light. The feature importance plot for a classifier trained to optimize the
redshift-weighted metric looks nearly identical.

\section{Results} \label{sec:results}

\subsection{Overall performance} \label{sec:overall_performance}

The results of both of our classifiers for many of the metrics defined in \ref{sec:metric} are shown in
Table~\ref{tab:metrics}. In general, we find that both classifiers have similar \changeda{performance} across all
of these global metrics. In the blinded phase of the PLAsTiCC, an earlier version of our algorithm won the
challenge with a score of 0.680 on the Kaggle metric (lower is better). The updated algorithm
presented in this paper achieves a slightly better score of 0.649 on this metric. This improvement mainly
came from restricting the allowable redshift range for the data augmentation procedure and propagating
uncertainties. Our original augmentation algorithm was allowed to modify the redshifts of objects arbitrarily.
At high redshifts, unreliable extrapolations of the GP models far into the restframe UV were being used to
produce the light curves. Additionally, we were not propagating the GP modeling uncertainties into the
generated fluxes. These issues were fixed for the version of the algorithm described in this paper which
dramatically improved performance at high redshifts.

\begin{deluxetable*}{lcc}
\tablecaption{Classifier performance on various metrics. \changeda{For the flat-weighted, redshift-weighted and Kaggle
metrics, lower numbers are better. An optimal classifier will have an AUC of 1, and higher AUCs are better.}
\label{tab:metrics}}
\tablehead{
    \colhead{Metric name} &
    \colhead{Flat-weighted classifier} &
    \colhead{Redshift-weighted classifier}
}
\startdata
Flat-weighted metric                            & 0.468 & 0.510 \\
Redshift-weighted metric                        & 0.523 & 0.500 \\
Kaggle metric                                   & 0.649 & 0.709 \\
\hline
AUC -- 90: Type Ia SN                           & 0.95721 & 0.95204 \\
AUC -- 67: Peculiar Type Ia SN -- 91bg-like     & 0.96672 & 0.96015 \\
AUC -- 52: Peculiar Type Ia SN -- SNIax         & 0.85988 & 0.84203 \\
AUC -- 42: Type II SN                           & 0.93570 & 0.90826 \\
AUC -- 62: Type Ibc SN                          & 0.92851 & 0.91558 \\
AUC -- 95: Superluminous SN (Magnetar model)    & 0.99442 & 0.99257 \\
AUC -- 15: Tidal disruption event               & 0.99254 & 0.99243 \\
AUC -- 64: Kilonova                             & 0.99815 & 0.99579 \\
AUC -- 88: Active galactic nuclei               & 0.99772 & 0.99706 \\
AUC -- 92: RR Lyrae                             & 0.99987 & 0.99986 \\
AUC -- 65: M-dwarf stellar flare                & 0.99999 & 0.99999 \\
AUC -- 16: Eclipsing binary stars               & 0.99983 & 0.99983 \\
AUC -- 53: Mira variables                       & 0.99947 & 0.99937 \\
AUC -- 6:  Microlens from single lens           & 0.99962 & 0.99966 \\
\enddata
\end{deluxetable*}

\begin{figure*}
    \plotone{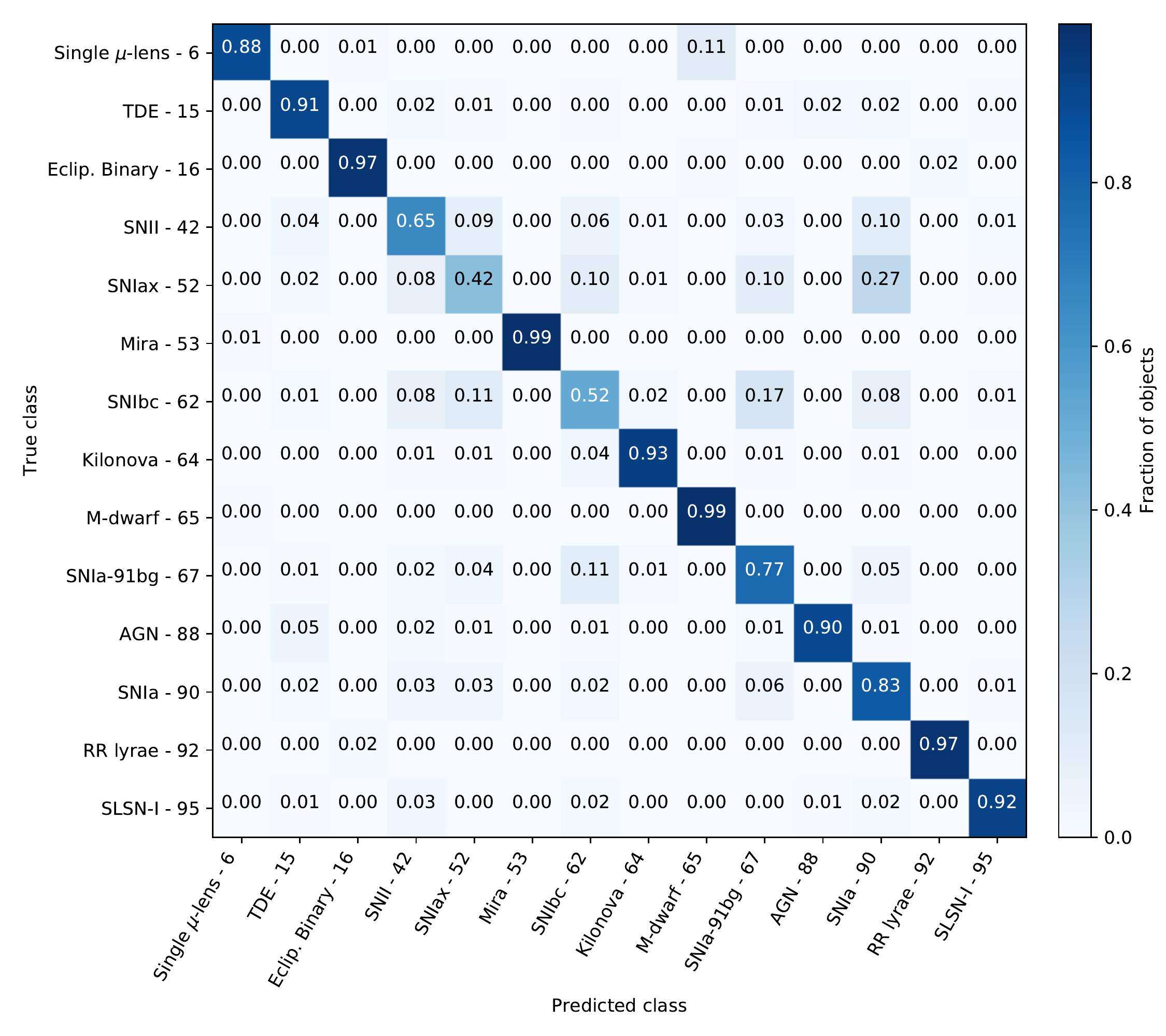}
    \caption{Confusion matrix for the flat-weighted classifier. We assign each object a ``predicted class''
    based \changeda{on} whichever class has the highest prediction probability for that object. We then calculate
    the fraction of objects for each true class that end up being predicted to be a given class. These fractions
    are shown in the figure for each pairing of true class and predicted class. A perfect classifier would have
    all ones for the diagonal terms, where the predicted class is the true class, and all zeros for the off-diagonal
    terms.}
    \label{fig:confusion_matrix}
\end{figure*}

Figure~\ref{fig:confusion_matrix} shows a confusion matrix for the flat-weighted classifier which can be used
to evaluate its performance on specific classes. For most classes, the top prediction of this
classifier is correct with an accuracy over over 80\%. The main challenge for the classifier is distinguishing
between the different types of supernovae. For example, Type~Iax supernovae are misclassified as Type~Ia
supernovae 27\% of the time, and Type~Ibc supernovae are misclassified as Type~Ia-91bg supernovae 17\% of the
time. This misclassification is often highly asymmetrical: only 3\% of Type~Ia supernovae are misclassified
as Type~Iax supernovae. This is likely due to the fact that the training set of Type~Iax supernovae (183 objects)
is small compared to the sample of Type~Ia supernovae (2,313 objects), and there is additional diversity in the
test set that is not seen in this small training set of Type~Iax supernovae. \changeda{Methods to address these
differences will be discussed in Section~\ref{sec:representativeness}. The redshift-weighted classifier
has a confusion matrix that is nearly identical \changedb{to the one} for the flat-weighted classifier. Note that the confusion matrix only considers
the top prediction for each object while our classifier outputs a probability for each object type, meaning
that there is additional information available for further analyses that is not captured by the confusion matrix.}

\subsection{Redshift-dependent performance} \label{sec:redshift_performance}

\begin{figure*}
    \epsscale{1.15}
    \plottwo{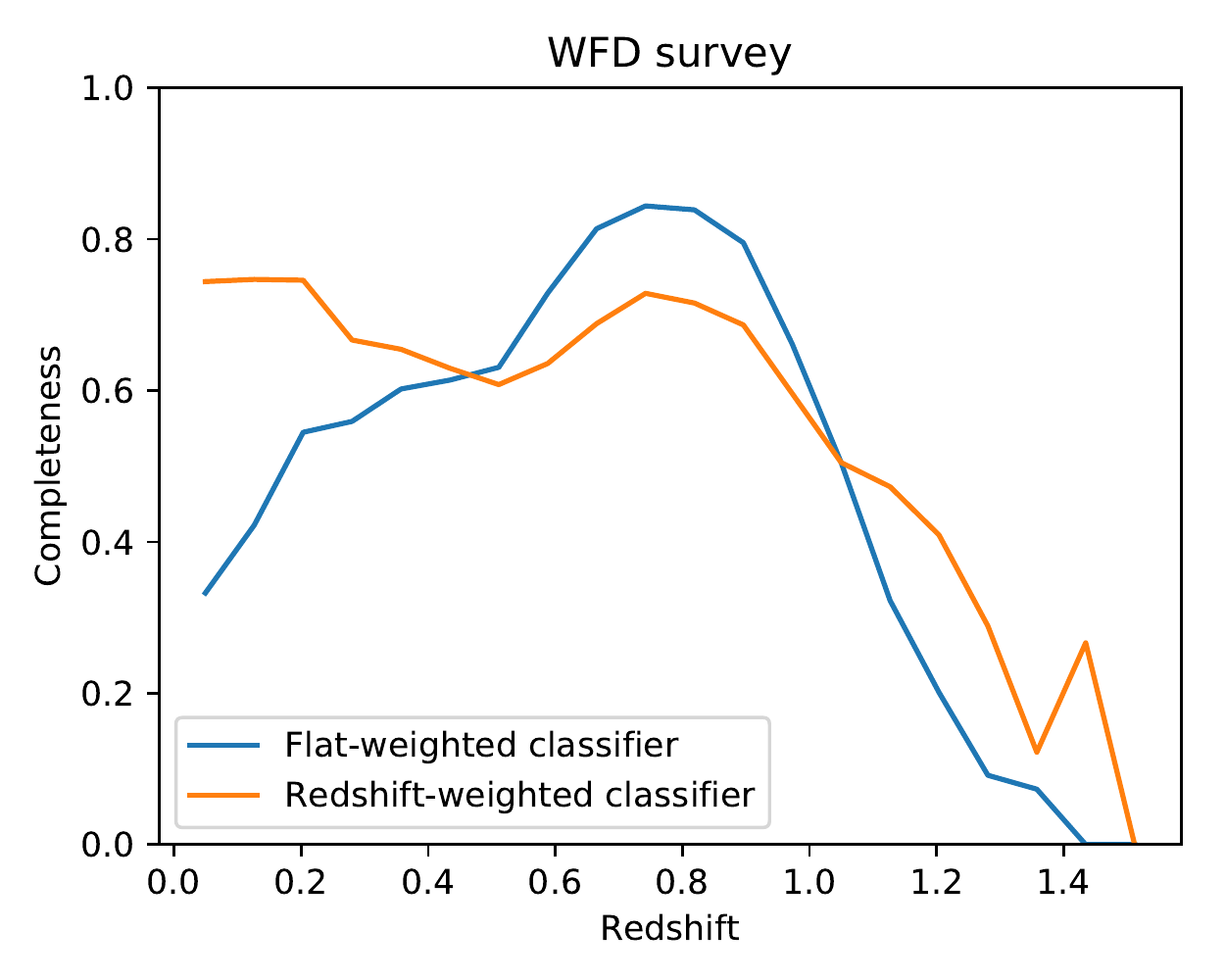}{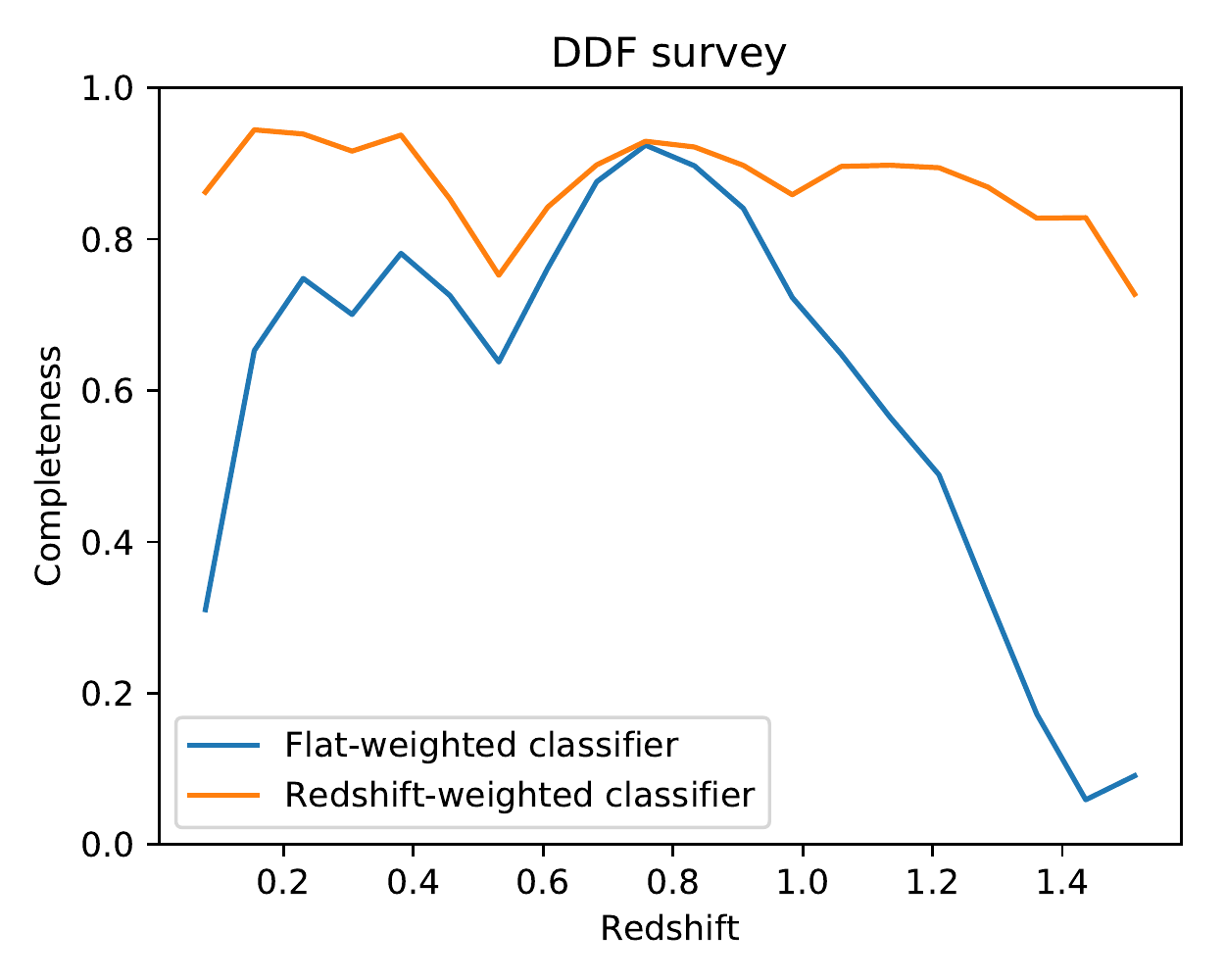}
    \caption{Completeness of the sample of SNe~Ia in the test set as a function of redshift for a fixed overall
    sample purity of 95\%. Left panel: Results for objects observed in the WFD survey. Right panel: Results for
    objects observed in the DDF survey. The Flat-weighted classifier produces classifications that encode the
    redshift distributions of different object types in the training set, and shows undesirable performance as a
    function of redshift. The redshift-weighted classifier produces classifications that are relatively stable
    with redshift. }
    \label{fig:completeness_redshift}
\end{figure*}

Despite the similarity in performance between the flat-weighted classifier and redshift-weighted classifier
on most global metrics, they exhibit very different \changeda{performance} as a function of redshift. 
For a fixed overall sample purity of 95\%, we calculate the completeness of the SN~Ia sample as a
function of redshift. The results of this procedure can be seen in Figure~\ref{fig:completeness_redshift}.

We find that the performance of the flat-weighted classifier has undesirable behavior as a function
of redshift. For objects observed in the WFD survey, we find that this classifier has its peak completeness
at redshift 0.8 where it correctly classifies 80\% of the SNe~Ia. At nearby redshifts, its
completeness drops to below 40\%. The reason for this strange performance can be seen in
Figure~\ref{fig:redshift_fractions}. Around redshift 0.8, nearly 70\% of the observed objects are Type
Ia supernovae, so the flat-weighted classifier is using information about the redshift distributions of the
different transients at different redshifts \changeda{as part of} its classification. Only 0.6\% of the SNe~Ia
in the sample are at a redshift less than 0.1, so the flat-weighted classifier infers that any object at
those redshifts is likely not to be a SN~Ia, and therefore has a very poor completeness for low-redshift SNe~Ia. We
see similar effects for the DDF survey.

Any classifier trained to optimize traditional metrics on a biased training sample will run into these issues if it
is allowed to incorporate redshift information into its predictions. Furthermore, the classifier is learning
\changeda{to use} the redshift distributions of objects in the training set, not the redshift distributions of ones in the test set. As
described in Section~\ref{sec:augmenting}, we expect there to be large discrepancies in the redshift distributions
of these two sets, even for augmented training sets. A classifier trained on the redshift-weighted metric provides
a solution to this problem by outputing a probability for each object type assuming that each class has the same
arbitrarily chosen redshift distribution in the training set. The performance of such a classifier will not depend
on the redshift distributions of the different kinds of objects in the training set. This can be seen in
Figure~\ref{fig:completeness_redshift}: for the WFD survey, the
redshift-weighted classifier shows a stable completeness of $\sim$70\% at low redshifts, and its completeness
declines nearly monotonically at higher redshifts as the signal-to-noise decreases. The redshift-weighted
classifier is able to classify objects in the DDF survey with a completeness of above 80\% at most redshifts,
maintaining this performance for even the highest-redshift SNe~Ia in the test sample at z=1.55.

\begin{figure*}
    \epsscale{1.15}
    \plottwo{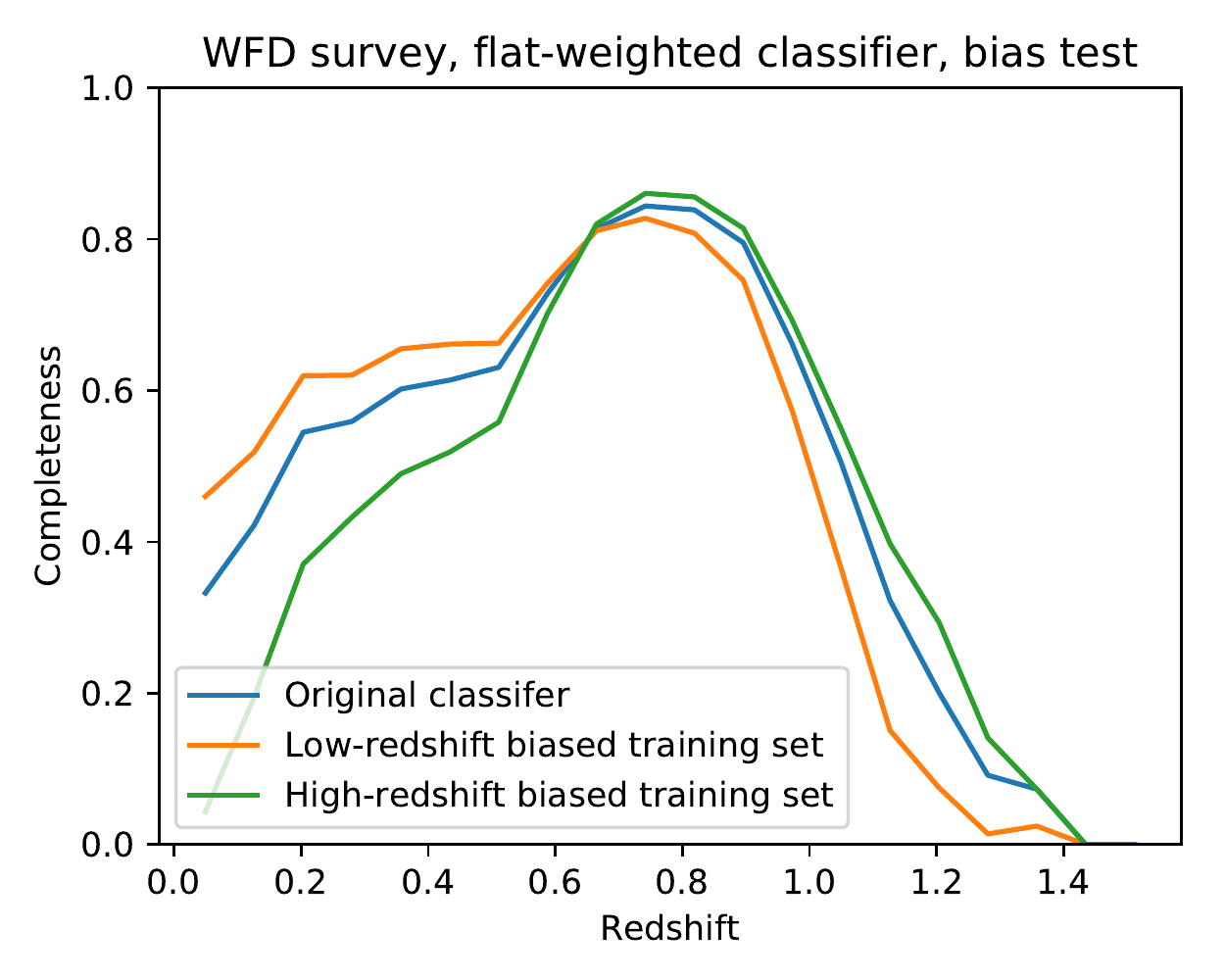}{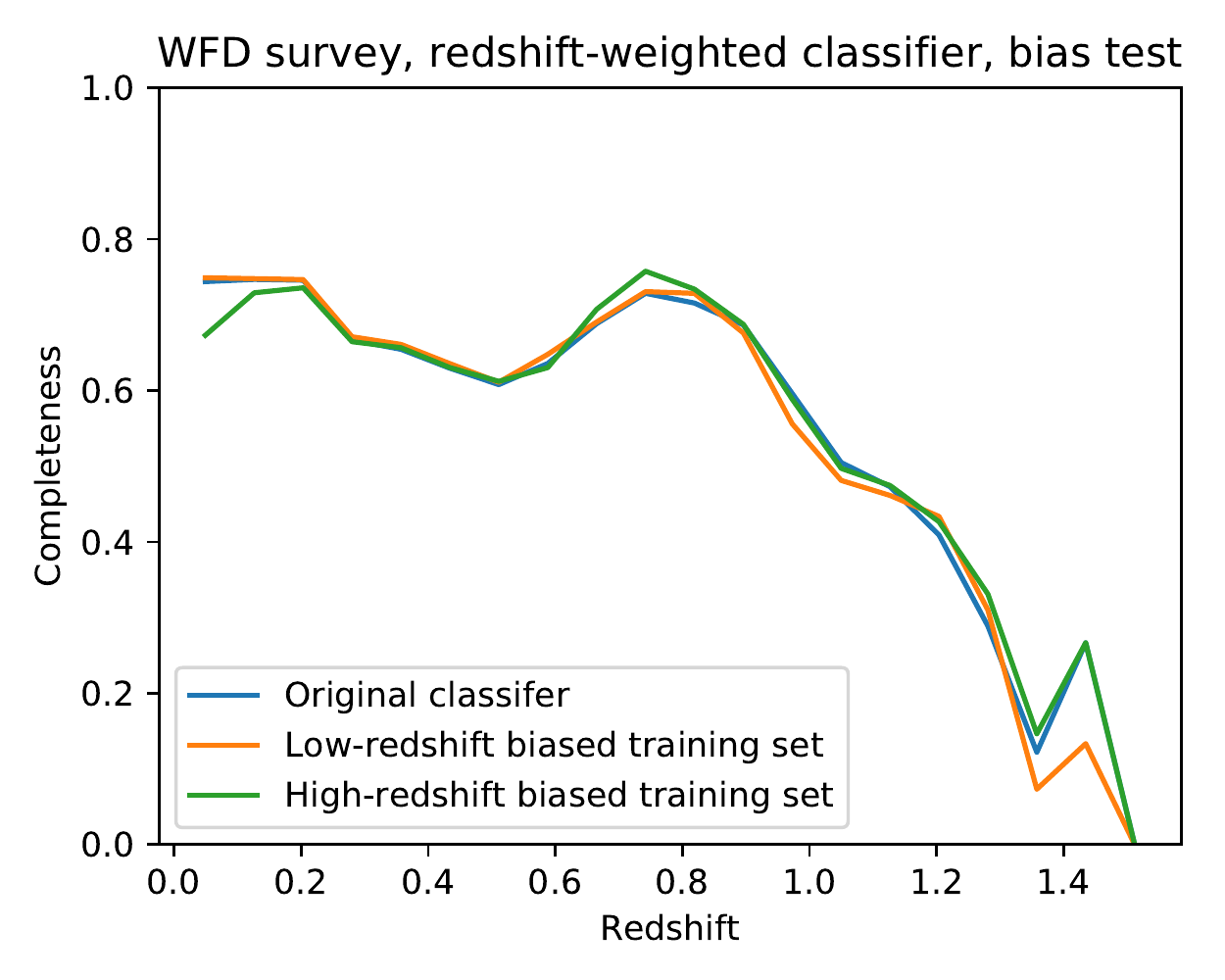}
    \caption{Completeness of the sample of SNe~Ia in the WFD sample of the test set as a function of redshift
    for a fixed overall sample purity of 95\% when different biases have been introduced in the training set.
    Left panel: Results for classifiers trained on the flat-weighted metric. Right panel: Results for classifiers
    trained on the redshift-weighted metric. The flat-weighted classifier is highly sensitive to the redshift
    distribution of objects in the training set while the redshift-weighted classifier shows very little difference
    in performance.}
    \label{fig:completeness_bias}
\end{figure*}

To validate the claim that the classifications produced by a classifier trained on the redshift-weighted metric
are independent of the redshift distributions of objects in the training set, we simulated modifying the
redshift distribution of the SNe~Ia in the training set. Starting with the same augmented training set used to
train the original classifiers, we create a \changeda{low-redshift-biased} training set by randomly dropping light curves of
SNe~Ia from the training set with probability $p(z) = \exp(-z)$. Similarly, we create a \changeda{high-redshift-biased}
training set by randomly dropping light curves of SNe~Ia from the training set with probability
$p(z) = min(\exp(z) - 1, 1)$. We keep all light curves from the transients that are not SNe~Ia so that only the
redshift distribution of SNe~Ia is changing between the original augmented training set and the biased training
sets. We then retrain our flat-weighted and redshift-weighted classifiers on these new training sets. The results of this
procedure can be seen in Figure~\ref{fig:completeness_bias}. For the flat-weighted classifier, the classifier
performance varies dramatically for these different training sets. At low and high redshifts, the difference
in completeness between the different classifiers varies by more than a factor of two. For the redshift-weighted
classifier, however, the different classifiers have nearly identical completenesses as a function of redshift.
Only small deviations from the original performance are seen at the very edges of the redshift range where
we have thrown out almost all of the SNe~Ia in the \changeda{biased training sets}.

\subsection{Comparison to other models}

A full comparison of our classifier to all of the other classifiers submitted to the blinded phase of the
PLAsTiCC will be presented in \plasticcresults. Those classifiers used a
wide range of techniques that were not explored in this paper, including template fitting, recurrent neural
networks and denoising autoencoders. However, all of those models were trained to optimize the Kaggle metric,
and we find that they exhibit the same problematic performance with redshift as discussed for our
flat-weighted classifier in Section~\ref{sec:redshift_performance}. The classifiers presented in this paper
are currently the best-performing models on the PLAsTiCC dataset, scoring 0.468 on the flat-weighted metric
compared to 0.481 for the next-best model submitted to the blinded phase of the PLAsTiCC challenge. On the
redshift-weighted metric, our redshift-weighted classifier scores 0.500, significantly better than the
next-best model which scores 0.609. Our classifier can therefore serve as a benchmark for future photometric
classifiers.

\section{Discussion} \label{sec:discussion}

\subsection{Computing resources} \label{sec:resources}

The main \changeda{limitation on computing speed} for this model is the GP regression. We perform our computation on an
Intel Xeon E3-1270 v3 CPU. A single core on this machine can fit the GP hyperparameters and extract features
for $\sim$10 objects per second. Training the LightGBM classifier takes $\sim$30 minutes for an augmented
training set of 591,410 objects after the features have been computed. Generating predictions with the trained
classifier takes a negligible amount of time compared to feature extraction ($\sim$1000 objects per second).
Processing the entire PLAsTiCC dataset therefore takes $\sim$100 core hours of computing time.

Predictions from this classifier can easily be done in real time. A machine with 100 CPU cores similar to
the one used for our testing could process $\sim$1000 objects per second and provide live typing estimates
for all of the transients and variables discovered in a survey. As new spectroscopic confirmations of objects
are obtained, the classifier could periodically be retrained to incorporate that new data into its training set.
The GP fits and feature extractions do not need to be redone for older data so long as the
feature extraction algorithm is unchanged, so retraining the classifier on new data can be done in under
an hour in most cases.

\subsection{Representativeness of the augmented training set} \label{sec:representativeness}

Ideally, our data augmentation procedure would produce an augmented training set that is fully representative of
the test set. In practice, however, there are several reasons why an augmented training set may differ from the test set.
First, the training set must cover the full intrinsic diversity of objects of the test set. Note that the
augmentation procedure
does not attempt to simulate new objects, it simply produces light curves for previously-measured objects under different
observing conditions and at different redshifts. If there are rare subtypes of objects that only appear in the test
set (e.g. peculiar supernovae), \changeda{and that have very different light curves from the objects in the training set},
then the augmentation procedure will not be able to produce light curves that
are similar to the ones observed for these objects. This issue can be addressed by using active learning
when obtaining the training set used for classification, as described in \citet{ishida19}. In this procedure,
the output of the classifier is used to determine which objects should be targeted for spectroscopic followup. This
helps to ensure that the original training set contains examples of all of the different object types that are
present in the full dataset.

The second major challenge for representativeness is \changeda{in} handling the different redshift distributions for object types
between the training and test sets. As discussed in Section~\ref{sec:augmenting}, the rates of different
transients are not currently known well enough as a function of redshift to produce an augmented dataset
that is truly representative of the test dataset. Instead, by training a classifier on a redshift-weighted metric,
we obtain a classifier whose output is independent of the redshift distributions of the classes in the training
sample, as shown in Section~\ref{sec:redshift_performance}. Such a classifier effectively produces classification
probabilities assuming the same arbitrarily chosen redshift distribution for every class. In our implementation,
we are using a log-uniform distribution in redshift for our classification, and this assumed distribution can easily
be propagated to further analyses. Analyses that depend on photometric classification, such as cosmology with SNe~Ia,
are already typically required to fit for or model the rates and selection efficiencies of different transient types as a
function of redshift to achieve their science goals. A redshift-weighted classifier produces output classification
probabilities that depend on known redshift distributions, and the biased redshift distributions in the original
training set will have \changeda{no effect on} the classification probabilities.

Our redshift-dependent classifier specifically addressed the issue of having different redshift distributions between
the training and test sets. A similar procedure could be applied to other observables of the transients, including
but not limited to their peak brightnesses, host properties, or rise and fall times. Assuming that these properties
can be measured accurately enough in the training set, we can simply reweight objects in the training set to force the
classifier to assume the same arbitrarily chosen distribution over this observable for each object type. This effectively
means that the classifier cannot \changeda{learn anything from} the distribution of this observable, or any indirect measurements of the
distribution of this observable to classify objects. The observable only needs to be available for the training set,
and does not necessarily need to be a feature that is used for classification. For example, our redshift-weighted
classifier is reweighted using the spectroscopic host redshifts \changeda{of} the objects in our training sample, but spectroscopic
host redshifts are not available for most objects in the full dataset. Nevertheless, the classifier outputs probabilities
that are independent of the distributions of spectroscopic host redshifts for different object types in the full dataset.

\subsection{Handling drifting subpopulations with redshift} \label{sec:redshift_drift}

One potential issue with any kind of photometric classification, including redshift augmentation, is drift in the
subpopulations of different types of transients as a function of redshift. For example, the properties of SNe~Ia are
correlated with the properties of their host galaxies such as host mass \citep{kelly10}. \changeda{As galaxy properties
evolve with redshift, the distributions of different subpopulations of SNe~Ia also evolve with redshift \citep{rigault13}.}
If the spectroscopically-confirmed training set is biased to low redshifts, then a
redshift-augmented training sample will have a different subpopulation distribution than the true one at high redshifts
leading to a bias in the classification probabilities. This is a challenge for any photometric classification method,
not just ones that use redshift augmentation, because the high-redshift followup strategy itself could be biased towards
some subpopulation.

\changedb{If the subpopulations} of an object type can be identified in the training set using some observable,
then we can reweight the training set as described previously to produce a classification assuming an arbitrarily chosen
distribution over this observable, with the same assumed distribution for all object types in the sample.
\changeda{For instance,} a dedicated
campaign to measure the host masses of every galaxy in the training set could be used to produce a photometric
classifier that classifies SNe~Ia without taking the host mass distributions, or any indirect measurements of the
host mass distributions, into account. This classifier will therefore produce classification probabilities that are
independent of the change in subpopulations as a function of redshift \changeda{associated with host mass}.
\changedb{Note that this procedure could introduce biases if the observable is not available for all objects in the
training sample, e.g. if it were not possible to reliably determine the host masses for higher redshift objects in the
training sample.}

\changedb{A final concern about changing subpopulations would be if new subpopulations appear at high 
redshifts that are not present at low redshifts. While the survey is running, active learning, as described in \citet{ishida19}
can be used to attempt to identify these subpopulations and trigger spectroscopic followup to add them to the training set.
If, however, these new subpopulations are missed entirely in the training set, then the classifier is unlikely to be able
to accurately classify objects from them.
}

\changedb{
Specific science applications may be more or less sensitive to the evolution of subpopulations with redshift.
For cosmology with SNe~Ia, for example, if a new subtype of SN~Ia appeared at high redshift that wasn't identified at
low redshift, then it could significantly bias the cosmological analysis. This is an issue regardless of the analysis
strategy: at high redshifts, a ``representative sample'' will likely only be able to spectroscopically type a small number
of objects, potentially missing \changedd{new} subpopulations. A lower-redshift followup strategy that intends to use
redshift-augmentation may be able to obtain a larger and more complete sample at the lower redshift, but it relies on
the assumption that we can produce a reasonable model of the differences between the low and high-redshift samples.
In practice, there is a trade-off between all of these concerns, and a variety of different followup strategies
should be simulated to determine the optimal strategy for the science objectives.}

\subsection{Implications for future surveys}

The results of this paper have several implications for future surveys. For surveys with limited spectroscopic
resources \changedb{devoted to obtaining training samples for photometric classification}, previous work (e.g.
\citetalias{lochner16}) has suggested that these surveys should attempt to produce
a training set of objects with spectroscopically-confirmed types that is as representative of the full dataset
as possible. With our augmentation technique, we instead suggest that spectroscopic resources may better be used to
obtain spectroscopic classifications of as many intermediate-redshift well-sampled light curves as possible.
We can then use these well-sampled light curves to simulate high-redshift light curves rather than
having to spend large amounts of spectroscopic followup to \changeda{type high-redshift objects} directly.

With augmentation, a light curve is most valuable if it has good-quality observations with a high cadence.
While the labels for all of the objects in the test set that the classifier will be applied to are not known, we
do know the cadence, observation depths, whether or not the light curve was impacted by season boundaries, etc.
of every object in the test set. A well sampled light curve can always be degraded to simulate all of these effects,
as shown in our augmentation procedure. One potential limitation of redshift augmentation is that the redshift range
to which a light curve model can be shifted is limited by the wavelength coverage of its input observations.
As discussed in Appendix~\ref{sec:appendix_augment}, we find that we must limit the change in wavelength to less
than 50\% to avoid having large GP extrapolation uncertainties for the bluer bands. This can be somewhat addressed
by adding additional followup in bluer bands, even including observations from other telescopes, but it effectively
means that there is a maximum change in redshift that can be applied in the augmentation procedure. For this reason,
intermediate-redshift light curves that can be augmented to high redshifts are more valuable to have in the training
set than very low-redshift light curves.

This suggests that an alternative strategy can be used to generate training sets for photometric classification.
At the start of a large survey such as LSST, a small, coordinated campaign can be undertaken to obtain very deep,
high-cadence observations in a limited region of the sky. Coincident spectroscopic campaigns can obtain redshifts
for many of the transients discovered in this \changeda{small survey.} These
light curves can then be used as templates for the augmentation procedure to produce light curves at any redshift,
and the well-observed light curves can easily be degraded to match the lower signal-to-noise and cadence light
curves from other parts of the survey. At this point, the classifier should be able to accurately classify the
majority of ``normal'' objects in the test sample.

\changedd{For unusual objects that are not present in this first
training set, we can attempt to use active learning following a procedure similar to \citet{ishida19} to identify
these objects and launch additional followup campaigns to obtain good quality observations for them. We can also
develop additional spectroscopic followup strategies tailored to the goals of specific scientific programs where
the presence of
unusual objects would impact the scientific results, such as searching for new subtypes of SNe~Ia at high
redshifts that could bias cosmological measurements.}

\changedb{Finally, spectroscopic observations will be obtained for purposes other than simply building training
sets for photometric classification or checking for new subpopulations of SNe~Ia at high redshifts. \changedd{For 
example,
there may be spectroscopic campaigns dedicated to specific subtypes or rare subpopulations of other object types.}
The teams obtaining these observations may have differing goals and selection requirements. It is essential to
coordinate spectroscopic efforts with other teams to best utilize the available spectroscopic resources.
}

\subsection{Improvements to the classifier}

There are several improvements that could be made to our classifier. First, for this analysis, we restricted ourselves
to a single GP kernel that was required to fit all the different kinds of
transients. Different kernels could perform better for different transients, so a natural extension is to
investigate the use of different kernels for the GP. As shown in \citetalias{revsbech18}, a
Gibbs kernel provides better fits to supernova-like light curves than the Mat\'ern kernel used for this
analysis. Periodic light curves, such as different types of variable stars, could be fit with periodic
kernels \citep{rasmussen06} to take advantage of the known periodicity. In general, each light curve
could be fit with GPs with multiple different kernels, and different features could be computed from each
of these fits, at the cost of increased computing time per object. The features from all of these different
fits could be used as input for a single LightGBM classifier. Additional features computed with other
means, such as features from a Lomb-Scargle periodogram \citep{vanderplas18} which have shown to be
very useful for classification of periodic light curves, could also easily be added to the classifier,
again, at the expense of additional computation time.

An additional approach for further work is to combine the results of multiple independent classifiers.
As described in \plasticcresults, combining the results of the top classifiers submitted to the
blinded phase of the PLAsTiCC resulted in much better performance than any single classifier.
These additional classifiers could also be trained on the augmented dataset to improve their performance.

\subsection{Conclusions}

We have developed a new framework for photometric classification of transients and variables from surveys
such as LSST. Our classifier is designed to be trained on datasets of spectroscopically-confirmed transients
and variables and is able to handle training sets that are biased towards bright low-redshift objects.
Using GP regression, we augment the set of light curves used for training to generate light curves over a wider
range of redshifts and observing conditions than present in the original training set. This procedure is
designed so that no specific model or parametrization of the light curve is required: it can be performed even
on poorly sampled light curves or ones with large gaps of observations in time.

Our classifier achieves the best performance on the PLAsTiCC dataset of any \changeda{single} algorithm to
date, scoring 0.468 on the metric defined by the PLAsTiCC team with flat weights for all objects in the training set.
It achieves an AUC of 0.957 for classification of SNe~Ia \changeda{compared to 0.934 for the next best single classifier
submitted to the blinded phase of the PLAsTiCC challenge. Our classifier sets a benchmark
for the performance of future photometric classifiers on the PLAsTiCC dataset and for the LSST.} Additionally, we
have shown how a metric
can be designed to produce classifiers whose output probabilities are independent of the redshift distributions
of the different kinds of transients in the training sample.  This leads to a better understanding of the output
probabilities of the classifier, which is essential for analyses
such as the determination of cosmological parameters using photometrically classified SNe~Ia.

All of the results in this paper can be reproduced with the \href{https://www.github.com/kboone/avocado}{\texttt{avocado}
classification package}. A Jupyter notebook is provided as part of that package which was used to produce all of
the figures shown in this paper.

\acknowledgments

We thank the PLAsTiCC team for producing this dataset and for organizing the PLAsTiCC. We thank the Kaggle
platform for hosting the blinded phase of the PLAsTiCC, and all of the teams who participated in this challenge
and shared ideas on the Kaggle discussion board. We appreciate the feedback that we received from Saul Perlmutter,
Greg Aldering, Ravi Gupta, Kara Ponder, Samantha Dixon, and Xiaosheng Huang. This work was supported by the
Director, Office of Science, Office of High Energy Physics of the U.S. Department of Energy under Contract
No. DE-AC02-05CH11231.

%



\software{
    Astropy \citep{astropy13, astropy18},
    George \citep{ambikasaran15},
    Jupyter \citep{kluyver16},
    LightGBM \citep{ke17},
    Matplotlib \citep{hunter07},
    NumPy \citep{vanderwalt11},
    Pandas \citep{mckinney10},
    scikit-learn \citep{scikit-learn},
    SciPy \citep{scipy}
}



\appendix

\section{Gaussian process regression} \label{appendix:gaussianprocess}

A stochastic process P(x) is a Gaussian process (GP) if for any finite set of points
${x_1, x_2, ..., x_n}$ the distribution ${P(x_1), P(x_2), ..., P(x_n)}$ is a multivariate normal
distribution. For the application of light curve modeling, a GP can be thought of as
a prior over a set of functions. By conditioning the GP on observations, we obtain a
posterior containing the set of functions that are consistent with the observations. For a detailed
discussion of GPs and their applications, see \citet{rasmussen06}.

A GP is uniquely defined by its mean function:
\begin{equation}
    \mu(x) = E[P(x)]
\end{equation}
and its covariance function, or ``kernel'':
\begin{equation}
    K(x_1, x_2) = E[(P(x_1) - \mu(x_1)) \times (P(x_2) - \mu(x_2))]
\end{equation}

The choice of $\mu$ and $K$ determines how different functions are weighted in the prior of the GP.
For this analysis, we choose the mean function of the GP to be zero so that the light curve
is modeled as a perturbation to an otherwise flat background. There are several possible choices for
the kernel. One common choice is the squared-exponential kernel (e.g. \citet{kim13} and
\citetalias{lochner16}). This kernel produces infinitely differentiable functions which can often be
unrealistically smooth and produce poor models of the data \citep{stein99}. In the context of the
PLAsTiCC dataset, objects such as the various explosive transients have sudden changes in their
light curves that will be smeared out by a squared-exponential kernel. For this reason, we choose instead
to use a Mat\'ern 3/2 kernel whose predictions are only once differentiable:
\begin{equation}
    K_{3/2}(x_1, x_2; \alpha, l) = \alpha^2 \left(1 + \sqrt{3 \frac{(x_1 - x_2)^2}{l^2}}\right) \exp\left(-\sqrt{3 \frac{(x_1 - x_2)^2}{l^2}}\right)
\end{equation}
The parameter $\alpha$ describes the amplitude scale of the functions that will be produced
by the GP, and $l$ sets the length scale over which functions vary.
Although we prefer the Mat\'ern 3/2 kernel for theoretical reasons, we do not notice significant
differences in the performance of the GP with this kernel compared to the squared-exponential one.


Most previous applications of GPs to astronomical light curves have modeled the different
bands independently. However, there are strong correlations in the light curve behavior between
different bands. Surveys such as LSST will not observe each band every night, so incorporating
cross-band information into the model is essential. As shown in \citet{fakhouri15}, a GP with a
two-dimensional kernel can be used with separate length scales in both time and in wavelength to
naturally incorporate this information. Our final GP kernel is the product of Mat\'ern kernels in
both wavelength and time:
\begin{equation}
    K_{2D}(t_1, t_2, \lambda_1, \lambda_2; \alpha, l_t, l_\lambda) = 
    \alpha^2 K_{3/2}(t_1, t_2; 1, l_t) K_{3/2}(\lambda_1, \lambda_2; 1, l_\lambda)
\end{equation}
This final kernel has three hyperparameters: the amplitude ($\alpha$) and length scales in both time ($l_t$)
and wavelength ($l_\lambda$).

\section{Implementation of the training set augmentation}
\label{sec:appendix_augment}

An overview of the augmentation procedure can be found in Section~\ref{sec:augmenting}. In this section,
we provide the details necessary to reproduce our procedure. In general, we attempt to use as few tuned
parameters as possible and to use the available information in the full dataset whenever possible. For
each object in the training sample, we generate up to 100 new versions of that object under different
observing conditions and at different redshifts.

For each new augmented extragalactic object, we first choose a new redshift for that object.
We limit the new redshift so that \changeda{$0.95~z_{original} < z_{new} < 5~z_{original}$}. A hard lower bound is used
here to avoid making faint objects much brighter and having their new simulated observations be
dominated by large modeling uncertainties. As our training set is biased toward \changeda{bright, low-redshift objects,}
this does not limit the performance of the classifier at low redshifts. A loose upper bound is used to
prevent the augmentation procedure from repeatedly trying to generate objects that are very faint and
not able to be detected by
the instrument. We impose an additional upper bound on the redshift of $1 + z_{new} < 1.5~(1 + z_{old})$.
If the wavelength range is shifted too far, the GP is required to extrapolate far from where there is
available data, and modeling uncertainties dominate its prediction. For this reason, we limit the possible
shift in wavelength to 50\% which results in the previous inequality. We choose a new redshift with a
log-uniform distribution between the lower and upper redshift bounds described previously. For augmented
\changeda{Galactic} objects, we simply modify the brightness of the object by adding an offset in magnitudes drawn
from a Gaussian distribution with a mean of 0 and a standard deviation of 0.5~mag.

Once a new redshift or brightness has been chosen, we choose to simulate it either as part of the WFD
survey or the DDF survey. Light curves in the DDF survey are much better sampled than light curves in the WFD
survey, so \changeda{for WFD template light curves we only generate WFD light curves}. For DDF template light curves,
we randomly choose to generate either a WFD or a DDF light curve. We choose a target number of observations for the
new light curve using a distribution that roughly matches the distribution of the PLAsTiCC test dataset.
For the DDF survey, we choose
a target number of observations following a Gaussian distribution with a mean of 330 and a standard
deviation of 30. For the WFD survey, we find that the number of observations of each light curve can
be described reasonably well with a three component Gaussian mixture model, with probabilities for each
component of [0.05, 0.4, 0.55], means of [95, 115, 138] observations and standard deviations of [20, 8, 8]
observations. \changeda{These numbers were all determined empirically by tuning our model to
match the observed distribution in the test set}.

We then choose the observation times and bands of these new observations. Our goal for this procedure is
to generate light curves under a wide range of different observing conditions. To account for a change in
redshift, we stretch the light curve in time \changeda{to account for the time dilation due to the redshift
difference. When shifting an object to higher redshifts, we fill in the light curve to account for the fact
that there is a lower density of observations after applying the time dilation
compared to at lower redshifts.} This is done by adding additional observations
to the new light curve at the same times as existing observations in the template light curve in randomly
selected bands. We do not attempt to sample observations at new times because we find that when doing so the uncertainty
in the GP flux predictions leads to unrealistic light curves with large uncertainties. For example, for the
light curve shown in Figure~\ref{fig:risetime_counts}, the GP flux predictions for the rise of the light curve
are highly unrealistic and have large uncertainties. Our method ensures that new light curves generated
using this light curve as a template will only include fall-time data where the GP flux predictions are accurate.

We shift the dates of the light curve by up to 50 days either forward or backward. This shifting does not
affect our classifier, but will affect classifiers that use absolute date information. To create light curves
that are cut off by season boundaries, we randomly drop a block of observations with a width of 250 days.
Following this, we randomly drop observations so that we have no more than the target number
of observations chosen previously, and we drop at least 10\% of observations \changeda{to ensure that we
are introducing some variation in the augmented light curves}.

After these procedures, we have chosen a set of observation times and bands for the new light curve. We
compute the mean GP flux prediction at those times and at a set of wavelengths corresponding to the bands shifted
by the new redshift.
We choose to use the mean GP flux predictions rather than drawing flux predictions from the GP because our goal is not
to produce all light curves consistent with our data, but instead to provide a reliable interpolation
or extrapolation of the template observations. We include the uncertainty of the GP flux predictions in the
uncertainties for each new observation.

Using the previously assigned new redshift, we adjust the brightness of the new light curve. This is done
assuming a fiducial flat $\Lambda$CDM cosmology with $H_0 = 70~\textrm{km}/\textrm{s}/\textrm{Mpc}$
and $\Omega_m = 0.3$. We find that the classification results are not sensitive to the choice of fiducial
cosmology. We then add simulated observation noise to each of the light curves. To determine the appropriate noise
levels for each of the WFD and DDF surveys, we fit lognormal distributions to a random subset of the observations
in the test set and draw from those distributions for our new simulated observations. We add this noise
in quadrature with the uncertainties from the GP flux predictions.

We naively estimate a photometric redshift for each of our new objects. For each new object, we draw one reference
\changeda{object} from the full PLAsTiCC dataset that has a spectroscopic redshift for its host. We then calculate the
difference between its photometric and spectroscopic redshifts, and we add this difference to the spectroscopic
redshift of the new object to estimate its photometric redshift. This procedure is not a proper model of photometric
redshifts, but it does ensure that the augmented training set contains any issues present for the photometric
redshifts in the full dataset. Incorporating a proper photometric redshift model into the augmenting procedure
would improve performance.

Finally, we apply a selection model to the augmented light curve. The original PLAsTiCC observations have a
detection flag that is set in a non-deterministic way. We fit an error function to the observations from the
full dataset to predict the probability of detection as a function of signal-to-noise, and use this to
predict the probability that an observation is flagged as detected. As was done for the original dataset, we
then only keep objects that have at least two detections in their light curves. If the object is rejected, then
we repeatedly attempt to generate a new light curve from the same template at the same redshift until the generated
object passes the selection criteria. If more than ten attempts at generating a new light curve fail to pass the
selection criteria, the template is skipped and we move on to the next template. This failure is common, and
typically indicates that we chose too high of a redshift for the new light curve, and that the generated fluxes
are too faint to be detectable by the telescope. We explicitly try to generate light curves at these high redshifts
to ensure that our augmented set covers the full range of light curves that it is possible to detect.

\begin{figure}
    \includegraphics[width=.32\linewidth]{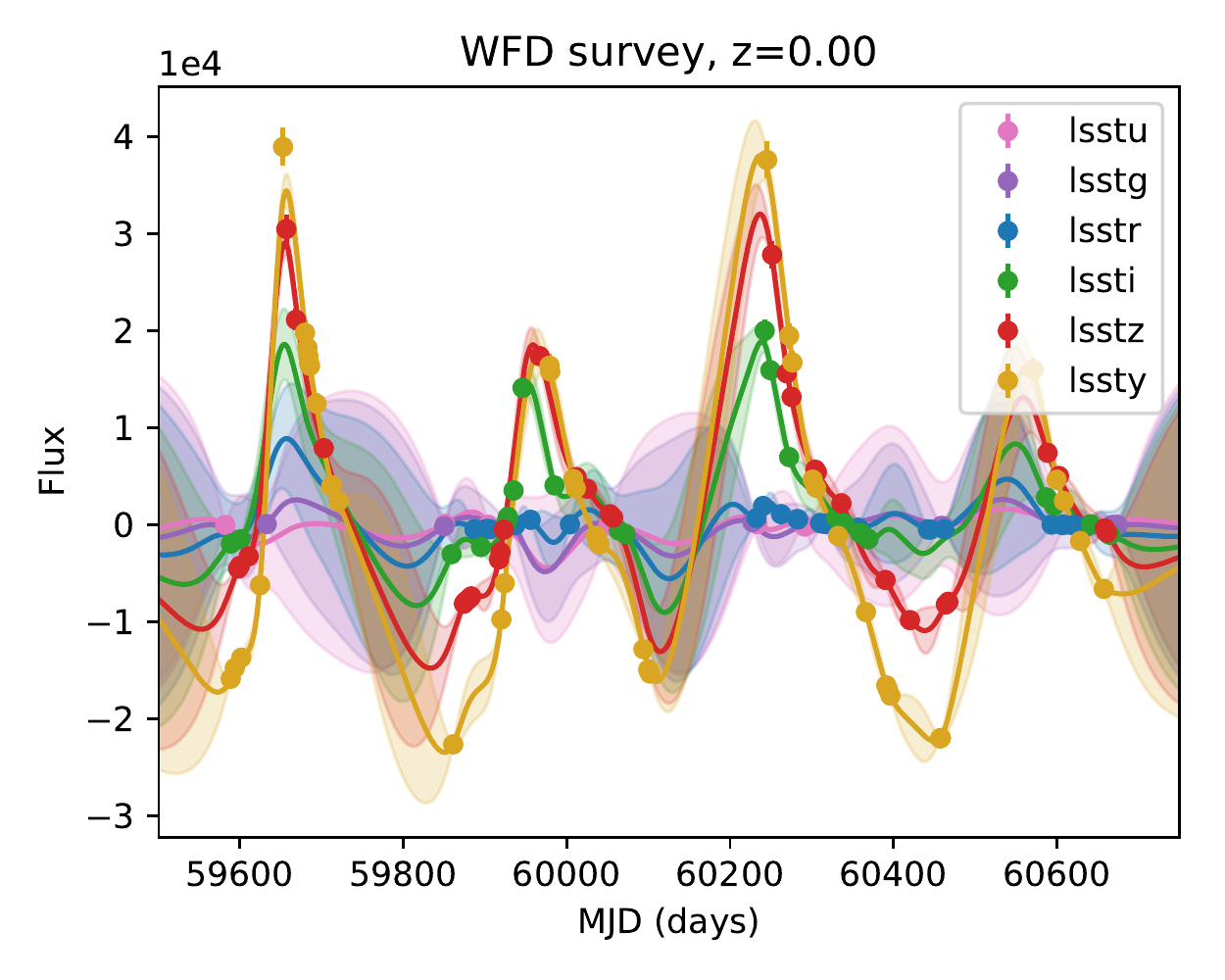}\quad\includegraphics[width=.32\linewidth]{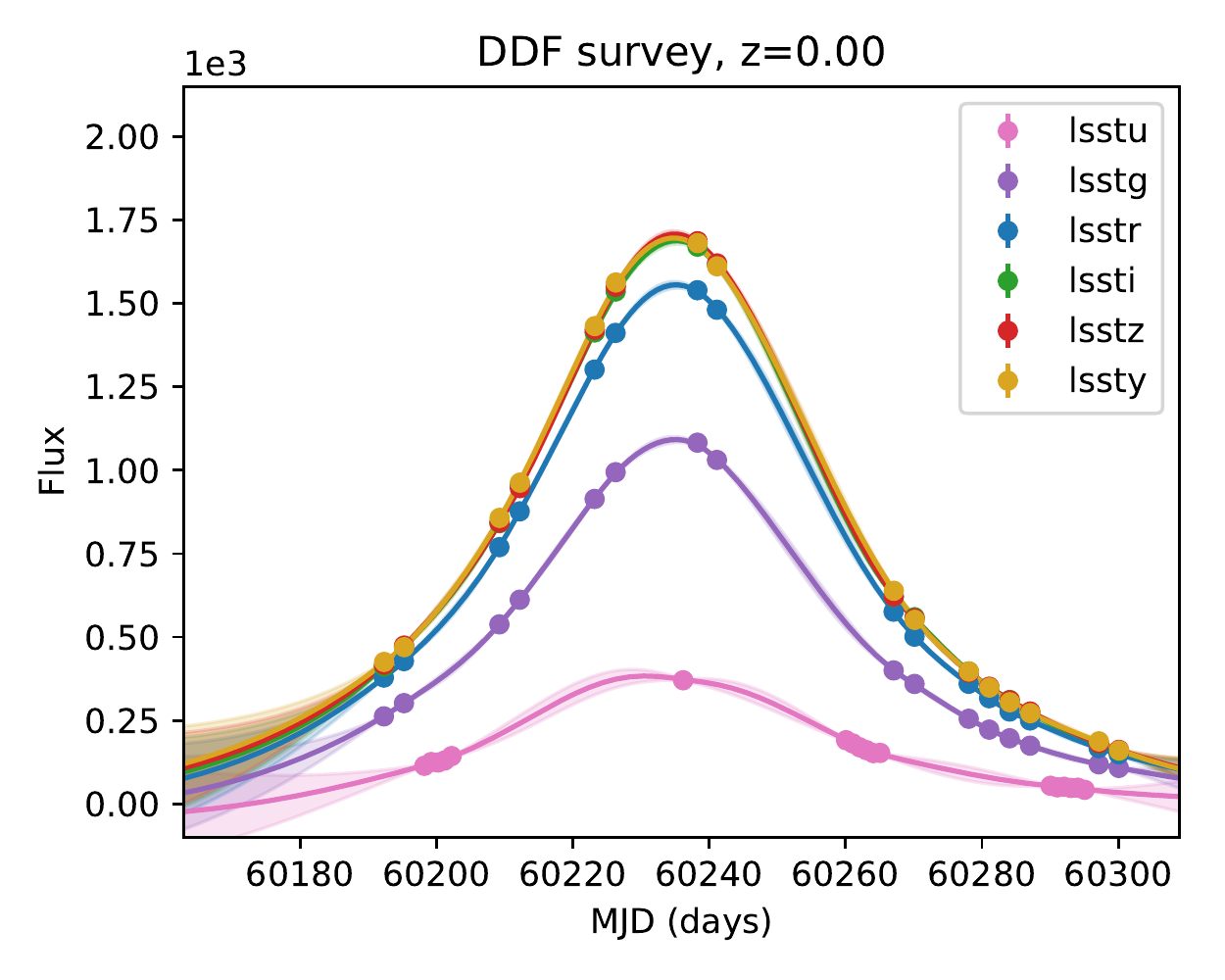}\quad\includegraphics[width=.32\linewidth]{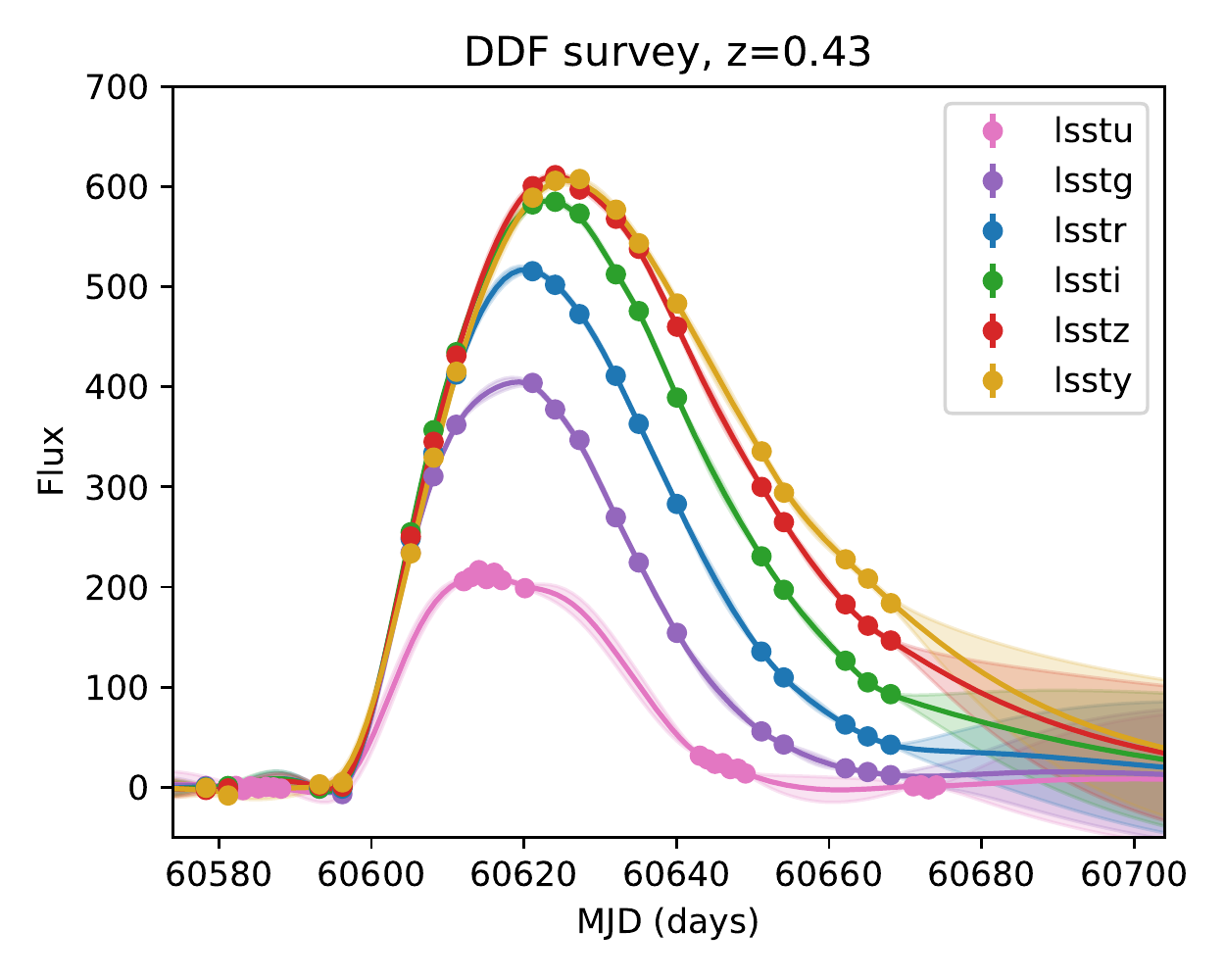}
    \\[\baselineskip]
    \includegraphics[width=.32\linewidth]{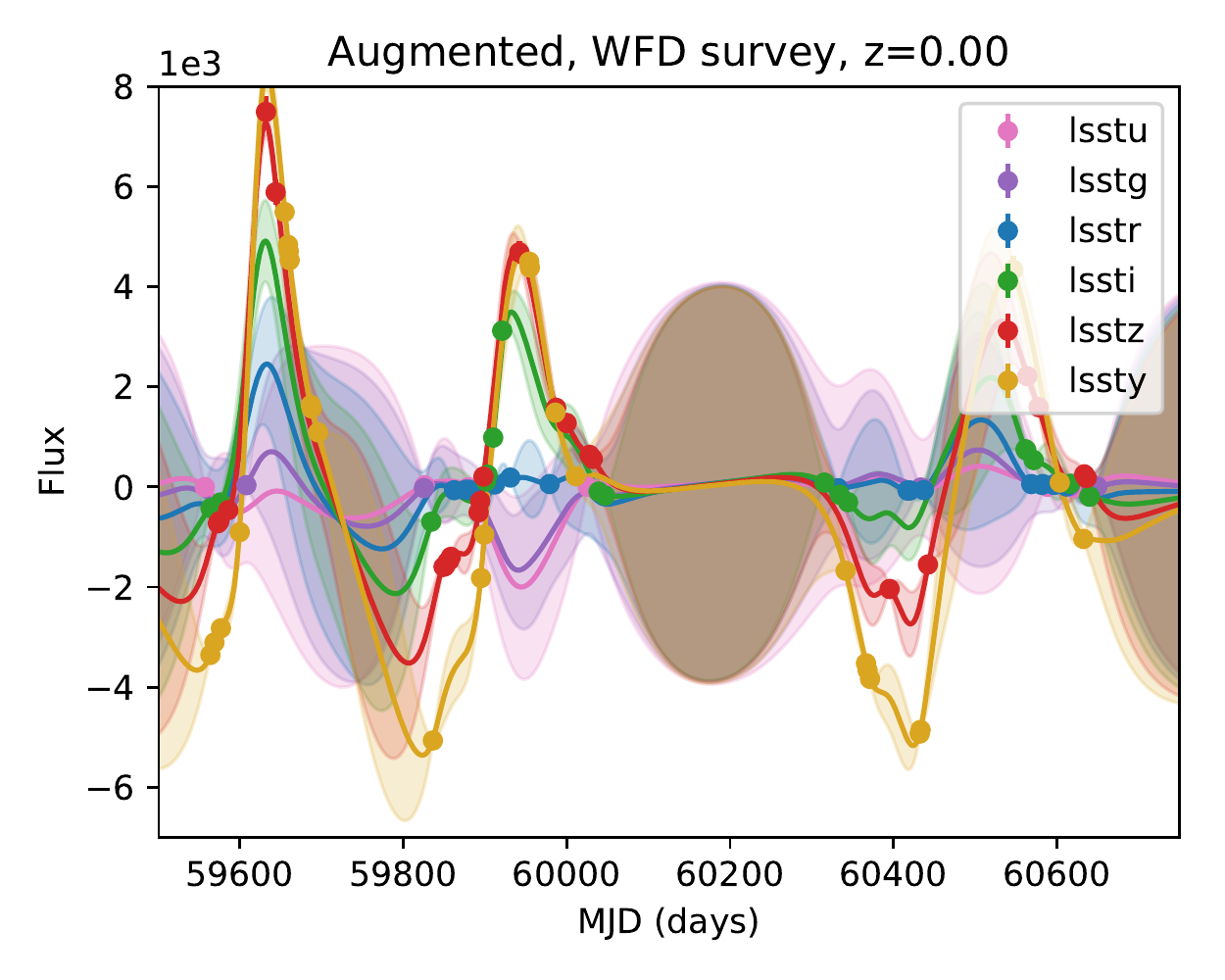}\quad\includegraphics[width=.32\linewidth]{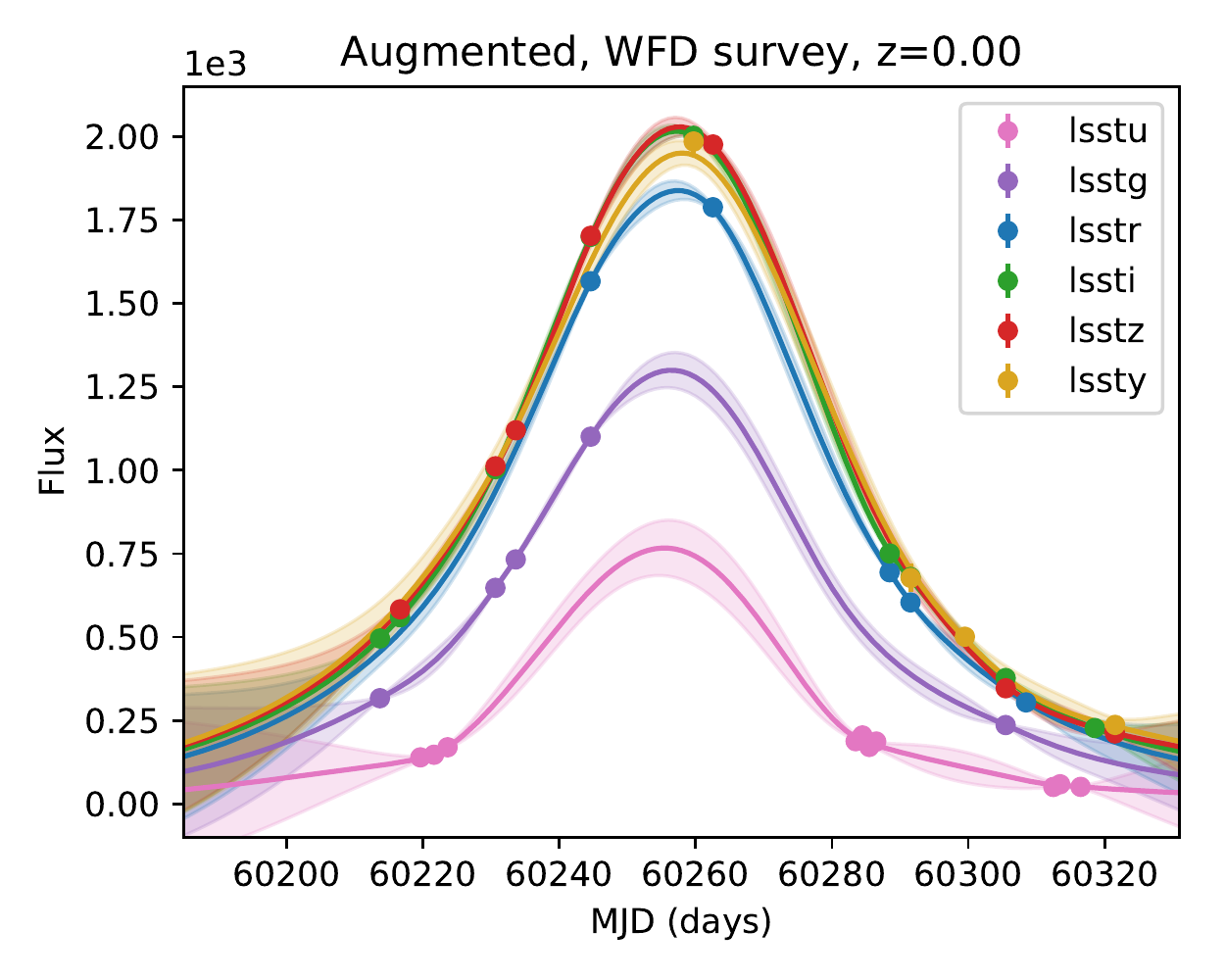}\quad\includegraphics[width=.32\linewidth]{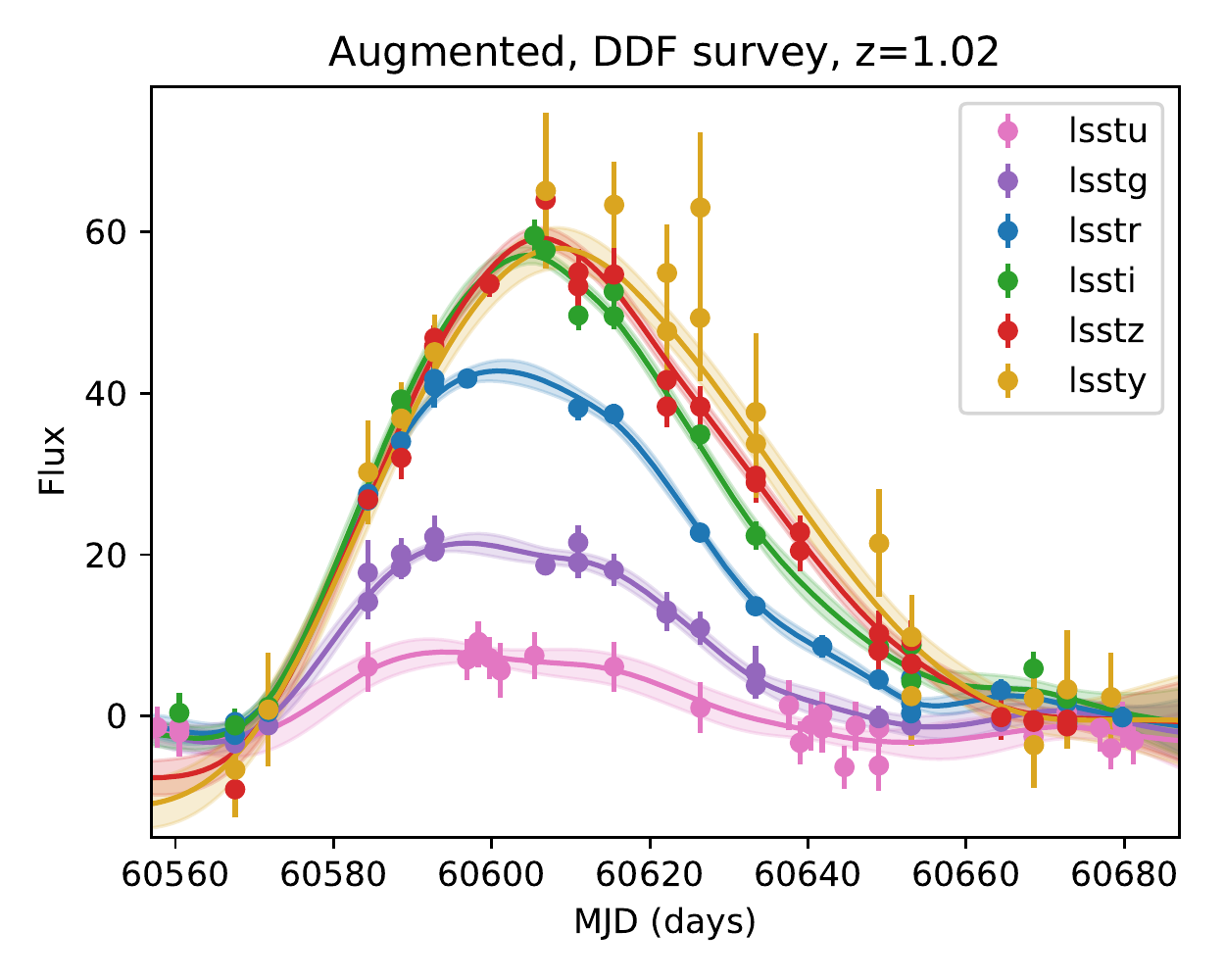}
    \\[\baselineskip]
    \includegraphics[width=.32\linewidth]{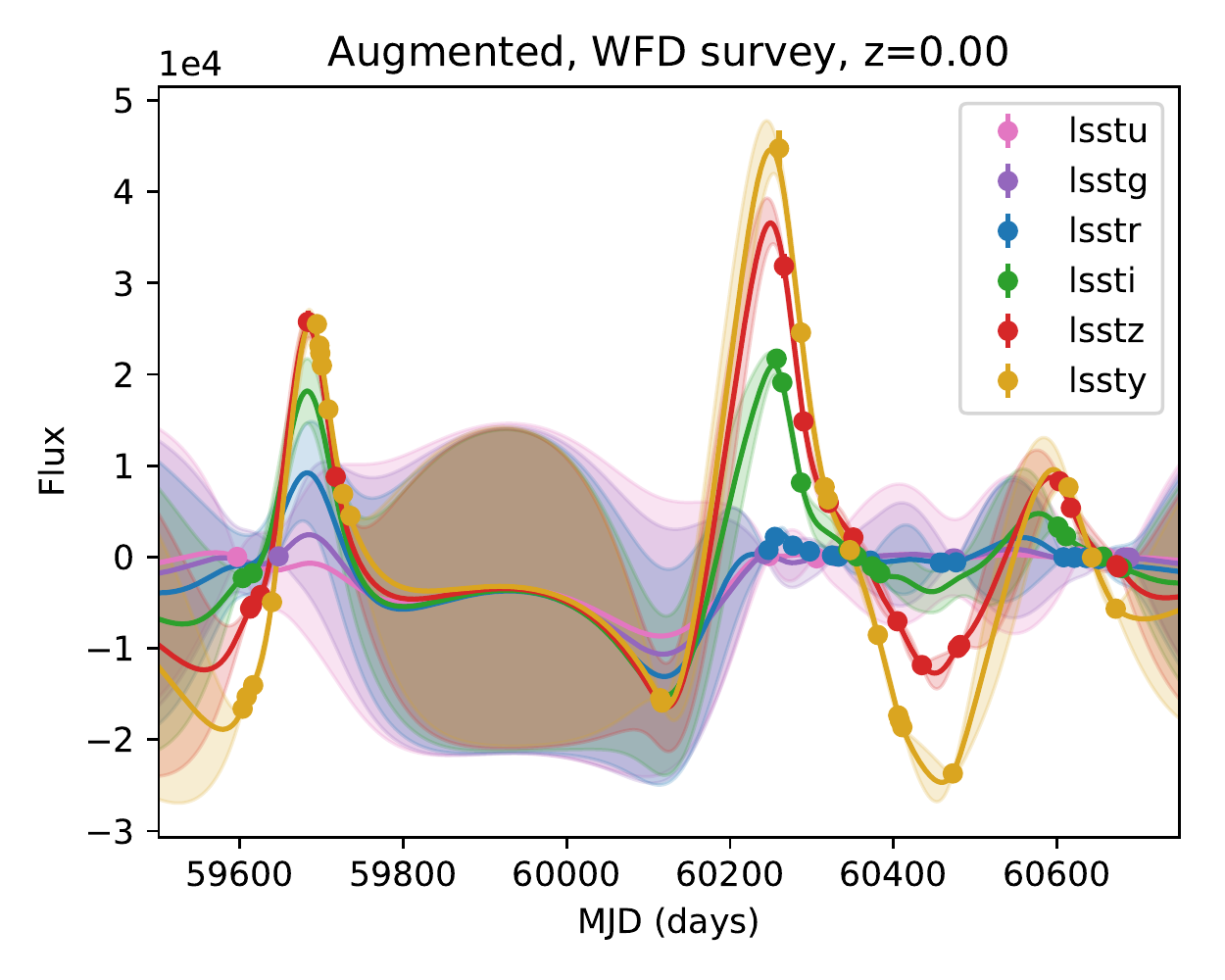}\quad\includegraphics[width=.32\linewidth]{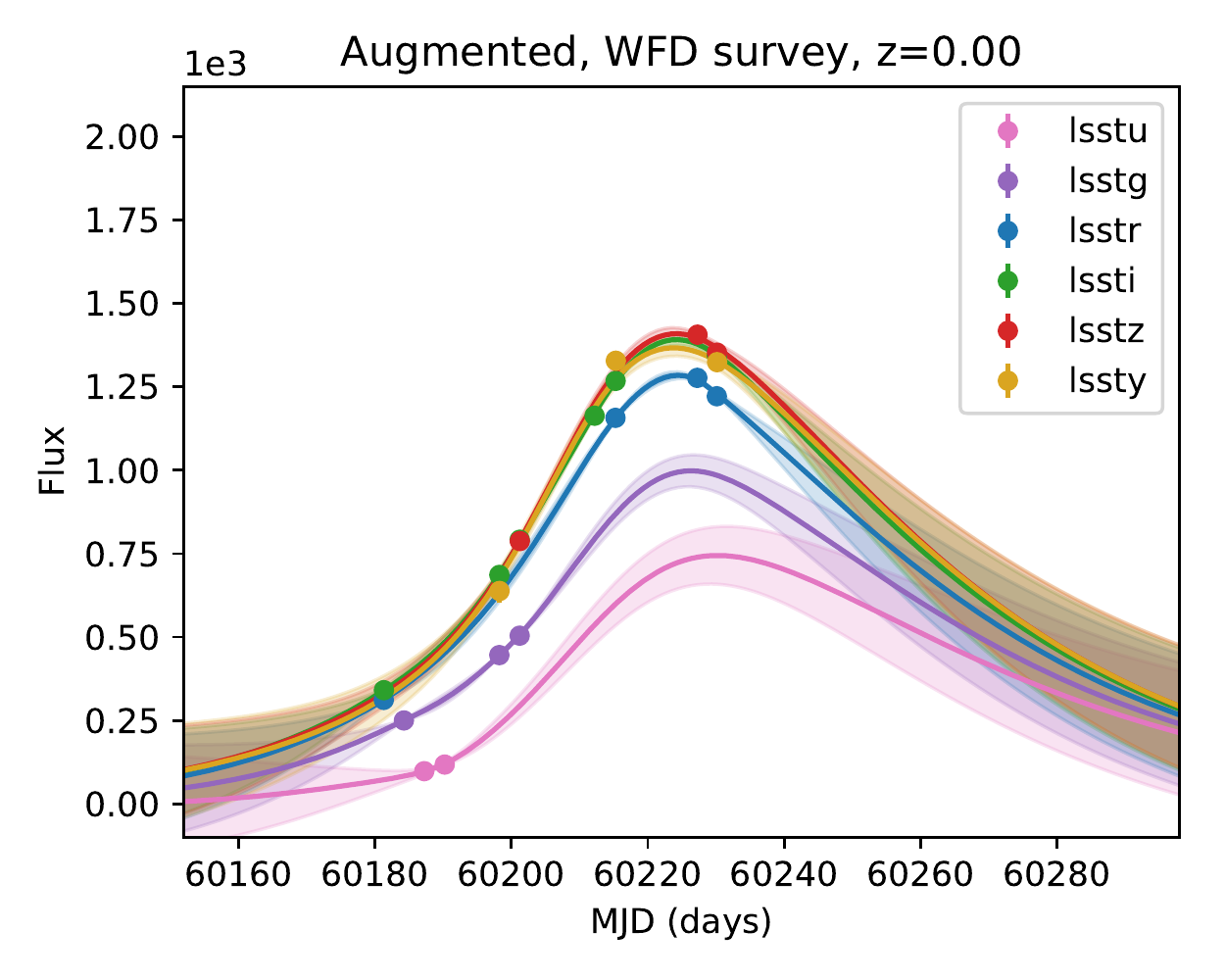}\quad\includegraphics[width=.32\linewidth]{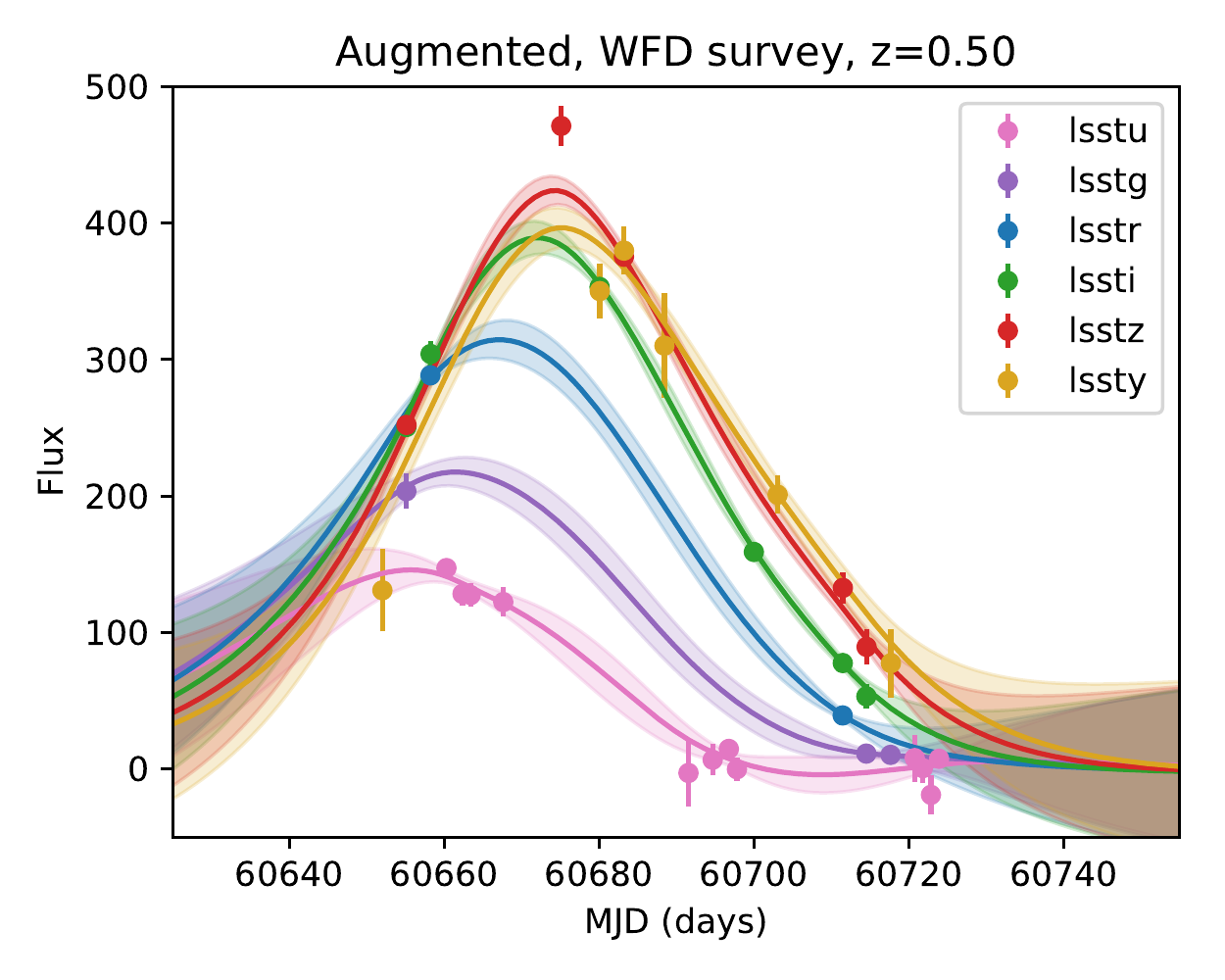}

    \caption{Examples of augmented light curves. Top row: original light curves. Bottom two rows:
    examples of light curves generated using the original light curves as a template. First column: example
    of a single Mira variable light curve. Second column: example of a single lens microlensing light curve.
    Third column: example of a superluminous supernova light curve. See Figure~\ref{fig:gp_snia} for an explanation
    of the lines and shading of the plots.}
    \label{fig:augmented_light_curves}
\end{figure}

Examples of the final augmented light curves are shown in Figure~\ref{fig:augmented_light_curves}.

\bibliography{avocado}

\begin{thebibliography}{}
\expandafter\ifx\csname natexlab\endcsname\relax\def\natexlab#1{#1}\fi
\providecommand{\url}[1]{\href{#1}{#1}}

\bibitem[{{Ambikasaran} {et~al.}(2015){Ambikasaran}, {Foreman-Mackey},
  {Greengard}, {Hogg}, \& {O'Neil}}]{ambikasaran15}
{Ambikasaran}, S., {Foreman-Mackey}, D., {Greengard}, L., {Hogg}, D.~W., \&
  {O'Neil}, M. 2015, IEEE Transactions on Pattern Analysis and Machine
  Intelligence, 38, 252

\bibitem[{{Astier} {et~al.}(2006){Astier}, {Guy}, {Regnault}, {Pain},
  {Aubourg}, {Balam}, {Basa}, {Carlberg}, {Fabbro}, {Fouchez}, {Hook},
  {Howell}, {Lafoux}, {Neill}, {Palanque-Delabrouille}, {Perrett}, {Pritchet},
  {Rich}, {Sullivan}, {Taillet}, {Aldering}, {Antilogus}, {Arsenijevic},
  {Balland}, {Baumont}, {Bronder}, {Courtois}, {Ellis}, {Filiol}, {Gon{\c
  c}alves}, {Goobar}, {Guide}, {Hardin}, {Lusset}, {Lidman}, {McMahon},
  {Mouchet}, {Mourao}, {Perlmutter}, {Ripoche}, {Tao}, \& {Walton}}]{astier06}
{Astier}, P., {Guy}, J., {Regnault}, N., {et~al.} 2006, \aap, 447, 31

\bibitem[{{Astropy Collaboration} {et~al.}(2013){Astropy Collaboration},
  {Robitaille}, {Tollerud}, {Greenfield}, {Droettboom}, {Bray}, {Aldcroft},
  {Davis}, {Ginsburg}, {Price-Whelan}, {Kerzendorf}, {Conley}, {Crighton},
  {Barbary}, {Muna}, {Ferguson}, {Grollier}, {Parikh}, {Nair}, {Unther},
  {Deil}, {Woillez}, {Conseil}, {Kramer}, {Turner}, {Singer}, {Fox}, {Weaver},
  {Zabalza}, {Edwards}, {Azalee Bostroem}, {Burke}, {Casey}, {Crawford},
  {Dencheva}, {Ely}, {Jenness}, {Labrie}, {Lim}, {Pierfederici}, {Pontzen},
  {Ptak}, {Refsdal}, {Servillat}, \& {Streicher}}]{astropy13}
{Astropy Collaboration}, {Robitaille}, T.~P., {Tollerud}, E.~J., {et~al.} 2013,
  \aap, 558, A33

\bibitem[{{Astropy Collaboration} {et~al.}(2018){Astropy Collaboration},
  {Price-Whelan}, {Sip{\H o}cz}, {G{\"u}nther}, {Lim}, {Crawford}, {Conseil},
  {Shupe}, {Craig}, {Dencheva}, {Ginsburg}, {VanderPlas}, {Bradley},
  {P{\'e}rez-Su{\'a}rez}, {de Val-Borro}, {Aldcroft}, {Cruz}, {Robitaille},
  {Tollerud}, {Ardelean}, {Babej}, {Bach}, {Bachetti}, {Bakanov}, {Bamford},
  {Barentsen}, {Barmby}, {Baumbach}, {Berry}, {Biscani}, {Boquien}, {Bostroem},
  {Bouma}, {Brammer}, {Bray}, {Breytenbach}, {Buddelmeijer}, {Burke},
  {Calderone}, {Cano Rodr{\'{\i}}guez}, {Cara}, {Cardoso}, {Cheedella},
  {Copin}, {Corrales}, {Crichton}, {D'Avella}, {Deil}, {Depagne}, {Dietrich},
  {Donath}, {Droettboom}, {Earl}, {Erben}, {Fabbro}, {Ferreira}, {Finethy},
  {Fox}, {Garrison}, {Gibbons}, {Goldstein}, {Gommers}, {Greco}, {Greenfield},
  {Groener}, {Grollier}, {Hagen}, {Hirst}, {Homeier}, {Horton}, {Hosseinzadeh},
  {Hu}, {Hunkeler}, {Ivezi{\'c}}, {Jain}, {Jenness}, {Kanarek}, {Kendrew},
  {Kern}, {Kerzendorf}, {Khvalko}, {King}, {Kirkby}, {Kulkarni}, {Kumar},
  {Lee}, {Lenz}, {Littlefair}, {Ma}, {Macleod}, {Mastropietro}, {McCully},
  {Montagnac}, {Morris}, {Mueller}, {Mumford}, {Muna}, {Murphy}, {Nelson},
  {Nguyen}, {Ninan}, {N{\"o}the}, {Ogaz}, {Oh}, {Parejko}, {Parley}, {Pascual},
  {Patil}, {Patil}, {Plunkett}, {Prochaska}, {Rastogi}, {Reddy Janga},
  {Sabater}, {Sakurikar}, {Seifert}, {Sherbert}, {Sherwood-Taylor}, {Shih},
  {Sick}, {Silbiger}, {Singanamalla}, {Singer}, {Sladen}, {Sooley},
  {Sornarajah}, {Streicher}, {Teuben}, {Thomas}, {Tremblay}, {Turner},
  {Terr{\'o}n}, {van Kerkwijk}, {de la Vega}, {Watkins}, {Weaver}, {Whitmore},
  {Woillez}, {Zabalza}, \& {Astropy Contributors}}]{astropy18}
{Astropy Collaboration}, {Price-Whelan}, A.~M., {Sip{\H o}cz}, B.~M., {et~al.}
  2018, \aj, 156, 123

\bibitem[{{Bailey} {et~al.}(2007){Bailey}, {Aragon}, {Romano}, {Thomas},
  {Weaver}, \& {Wong}}]{bailey07}
{Bailey}, S., {Aragon}, C., {Romano}, R., {et~al.} 2007, \apj, 665, 1246

\bibitem[{{Beers} {et~al.}(1990){Beers}, {Flynn}, \& {Gebhardt}}]{beers90}
{Beers}, T.~C., {Flynn}, K., \& {Gebhardt}, K. 1990, \aj, 100, 32

\bibitem[{{Bernstein} {et~al.}(2012){Bernstein}, {Kessler}, {Kuhlmann},
  {Biswas}, {Kovacs}, {Aldering}, {Crane}, {D'Andrea}, {Finley}, {Frieman},
  {Hufford}, {Jarvis}, {Kim}, {Marriner}, {Mukherjee}, {Nichol}, {Nugent},
  {Parkinson}, {Reis}, {Sako}, {Spinka}, \& {Sullivan}}]{bernstein12}
{Bernstein}, J.~P., {Kessler}, R., {Kuhlmann}, S., {et~al.} 2012, \apj, 753,
  152

\bibitem[{{Betoule} {et~al.}(2014){Betoule}, {Kessler}, {Guy}, {Mosher},
  {Hardin}, {Biswas}, {Astier}, {El-Hage}, {Konig}, {Kuhlmann}, {Marriner},
  {Pain}, {Regnault}, {Balland}, {Bassett}, {Brown}, {Campbell}, {Carlberg},
  {Cellier-Holzem}, {Cinabro}, {Conley}, {D'Andrea}, {DePoy}, {Doi}, {Ellis},
  {Fabbro}, {Filippenko}, {Foley}, {Frieman}, {Fouchez}, {Galbany}, {Goobar},
  {Gupta}, {Hill}, {Hlozek}, {Hogan}, {Hook}, {Howell}, {Jha}, {Le Guillou},
  {Leloudas}, {Lidman}, {Marshall}, {M{\"o}ller}, {Mour{\~a}o}, {Neveu},
  {Nichol}, {Olmstead}, {Palanque-Delabrouille}, {Perlmutter}, {Prieto},
  {Pritchet}, {Richmond}, {Riess}, {Ruhlmann-Kleider}, {Sako}, {Schahmaneche},
  {Schneider}, {Smith}, {Sollerman}, {Sullivan}, {Walton}, \&
  {Wheeler}}]{betoule14}
{Betoule}, M., {Kessler}, R., {Guy}, J., {et~al.} 2014, \aap, 568, A22

\bibitem[{{Charnock} \& {Moss}(2017)}]{charnock17}
{Charnock}, T., \& {Moss}, A. 2017, \apj, 837, L28

\bibitem[{{Delgado} {et~al.}(2014){Delgado}, {Saha}, {Chandrasekharan}, {Cook},
  {Petry}, \& {Ridgway}}]{delgado14}
{Delgado}, F., {Saha}, A., {Chandrasekharan}, S., {et~al.} 2014, in \procspie,
  Vol. 9150, Modeling, Systems Engineering, and Project Management for
  Astronomy VI, 915015

\bibitem[{{Fakhouri} {et~al.}(2015){Fakhouri}, {Boone}, {Aldering},
  {Antilogus}, {Aragon}, {Bailey}, {Baltay}, {Barbary}, {Baugh}, {Bongard},
  {Buton}, {Chen}, {Childress}, {Chotard}, {Copin}, {Fagrelius}, {Feindt},
  {Fleury}, {Fouchez}, {Gangler}, {Hayden}, {Kim}, {Kowalski}, {Leget},
  {Lombardo}, {Nordin}, {Pain}, {Pecontal}, {Pereira}, {Perlmutter},
  {Rabinowitz}, {Ren}, {Rigault}, {Rubin}, {Runge}, {Saunders}, {Scalzo},
  {Smadja}, {Sofiatti}, {Strovink}, {Suzuki}, {Tao}, {Thomas}, {Weaver}, \&
  {Nearby Supernova Factory}}]{fakhouri15}
{Fakhouri}, H.~K., {Boone}, K., {Aldering}, G., {et~al.} 2015, \apj, 815, 58

\bibitem[{{Guy} {et~al.}(2007){Guy}, {Astier}, {Baumont}, {Hardin}, {Pain},
  {Regnault}, {Basa}, {Carlberg}, {Conley}, {Fabbro}, {Fouchez}, {Hook},
  {Howell}, {Perrett}, {Pritchet}, {Rich}, {Sullivan}, {Antilogus}, {Aubourg},
  {Bazin}, {Bronder}, {Filiol}, {Palanque-Delabrouille}, {Ripoche}, \&
  {Ruhlmann-Kleider}}]{guy07}
{Guy}, J., {Astier}, P., {Baumont}, S., {et~al.} 2007, \aap, 466, 11

\bibitem[{{Hlozek} {et~al.}(2012){Hlozek}, {Kunz}, {Bassett}, {Smith},
  {Newling}, {Varughese}, {Kessler}, {Bernstein}, {Campbell}, {Dilday},
  {Falck}, {Frieman}, {Kuhlmann}, {Lampeitl}, {Marriner}, {Nichol}, {Riess},
  {Sako}, \& {Schneider}}]{hlozek12}
{Hlozek}, R., {Kunz}, M., {Bassett}, B., {et~al.} 2012, \apj, 752, 79

\bibitem[{Hunter(2007)}]{hunter07}
Hunter, J.~D. 2007, Computing in Science \& Engineering, 9, 90

\bibitem[{{Ishida} \& {de Souza}(2013)}]{ishida13}
{Ishida}, E.~E.~O., \& {de Souza}, R.~S. 2013, \mnras, 430, 509

\bibitem[{{Ishida} {et~al.}(2019){Ishida}, {Beck}, {Gonz{\'a}lez-Gait{\'a}n},
  {de Souza}, {Krone-Martins}, {Barrett}, {Kennamer}, {Vilalta}, {Burgess},
  {Quint}, {Vitorelli}, {Mahabal}, \& {Gangler}}]{ishida19}
{Ishida}, E.~E.~O., {Beck}, R., {Gonz{\'a}lez-Gait{\'a}n}, S., {et~al.} 2019,
  \mnras, 483, 2

\bibitem[{{Jones} {et~al.}(2017){Jones}, {Scolnic}, {Riess}, {Kessler}, {Rest},
  {Kirshner}, {Berger}, {Ortega}, {Foley}, {Chornock}, {Challis}, {Burgett},
  {Chambers}, {Draper}, {Flewelling}, {Huber}, {Kaiser}, {Kudritzki},
  {Metcalfe}, {Wainscoat}, \& {Waters}}]{jones17}
{Jones}, D.~O., {Scolnic}, D.~M., {Riess}, A.~G., {et~al.} 2017, \apj, 843, 6

\bibitem[{Jones {et~al.}(2001)Jones, Oliphant, Peterson, {et~al.}}]{scipy}
Jones, E., Oliphant, T., Peterson, P., {et~al.} 2001, {SciPy}: Open source
  scientific tools for {Python}, v1.2.1,  Online, [accessed 6/3/2019].
\newblock \url{http://www.scipy.org/}

\bibitem[{{Kaiser} {et~al.}(2010){Kaiser}, {Burgett}, {Chambers}, {Denneau},
  {Heasley}, {Jedicke}, {Magnier}, {Morgan}, {Onaka}, \& {Tonry}}]{kaiser10}
{Kaiser}, N., {Burgett}, W., {Chambers}, K., {et~al.} 2010, in \procspie, Vol.
  7733, Ground-based and Airborne Telescopes III, 77330E

\bibitem[{{Karpenka} {et~al.}(2013){Karpenka}, {Feroz}, \&
  {Hobson}}]{karpenka13}
{Karpenka}, N.~V., {Feroz}, F., \& {Hobson}, M.~P. 2013, \mnras, 429, 1278

\bibitem[{Ke {et~al.}(2017)Ke, Meng, Finley, Wang, Chen, Ma, Ye, \& Liu}]{ke17}
Ke, G., Meng, Q., Finley, T., {et~al.} 2017, in Advances in Neural Information
  Processing Systems 30, ed. I.~Guyon, U.~V. Luxburg, S.~Bengio, H.~Wallach,
  R.~Fergus, S.~Vishwanathan, \& R.~Garnett (Curran Associates, Inc.),
  3146--3154

\bibitem[{{Kelly} {et~al.}(2010){Kelly}, {Hicken}, {Burke}, {Mandel}, \&
  {Kirshner}}]{kelly10}
{Kelly}, P.~L., {Hicken}, M., {Burke}, D.~L., {Mandel}, K.~S., \& {Kirshner},
  R.~P. 2010, \apj, 715, 743

\bibitem[{{Kessler} {et~al.}(2009){Kessler}, {Bernstein}, {Cinabro}, {Dilday},
  {Frieman}, {Jha}, {Kuhlmann}, {Miknaitis}, {Sako}, {Taylor}, \&
  {Vanderplas}}]{kessler09}
{Kessler}, R., {Bernstein}, J.~P., {Cinabro}, D., {et~al.} 2009, \pasp, 121,
  1028

\bibitem[{{Kessler} {et~al.}(2010){Kessler}, {Bassett}, {Belov}, {Bhatnagar},
  {Campbell}, {Conley}, {Frieman}, {Glazov}, {Gonz{\'a}lez-Gait{\'a}n},
  {Hlozek}, {Jha}, {Kuhlmann}, {Kunz}, {Lampeitl}, {Mahabal}, {Newling},
  {Nichol}, {Parkinson}, {Sajeeth Philip}, {Poznanski}, {Richards}, {Rodney},
  {Sako}, {Schneider}, {Smith}, {Stritzinger}, \& {Varughese}}]{kessler10}
{Kessler}, R., {Bassett}, B., {Belov}, P., {et~al.} 2010, \pasp, 122, 1415

\bibitem[{{Kessler} {et~al.}(2015){Kessler}, {Marriner}, {Childress},
  {Covarrubias}, {D'Andrea}, {Finley}, {Fischer}, {Foley}, {Goldstein},
  {Gupta}, {Kuehn}, {Marcha}, {Nichol}, {Papadopoulos}, {Sako}, {Scolnic},
  {Smith}, {Sullivan}, {Wester}, {Yuan}, {Abbott}, {Abdalla}, {Allam},
  {Benoit-L{\'e}vy}, {Bernstein}, {Bertin}, {Brooks}, {Carnero Rosell},
  {Carrasco Kind}, {Castander}, {Crocce}, {da Costa}, {Desai}, {Diehl},
  {Eifler}, {Fausti Neto}, {Flaugher}, {Frieman}, {Gerdes}, {Gruen}, {Gruendl},
  {Honscheid}, {James}, {Kuropatkin}, {Li}, {Maia}, {Marshall}, {Martini},
  {Miller}, {Miquel}, {Nord}, {Ogando}, {Plazas}, {Reil}, {Romer}, {Roodman},
  {Sanchez}, {Sevilla-Noarbe}, {Smith}, {Soares-Santos}, {Sobreira}, {Tarle},
  {Thaler}, {Thomas}, {Tucker}, {Walker}, \& {DES Collaboration}}]{kessler15}
{Kessler}, R., {Marriner}, J., {Childress}, M., {et~al.} 2015, \aj, 150, 172

\bibitem[{{Kessler} {et~al.}(2019){Kessler}, {Narayan}, {Avelino}, {Bachelet},
  {Biswas}, {Brown}, {Chernoff}, {Connolly}, {Dai}, {Daniel}, {Di Stefano},
  {Drout}, {Galbany}, {Gonz{\'a}lez-Gait{\'a}n}, {Graham}, {Hlo{\v z}ek},
  {Ishida}, {Guillochon}, {Jha}, {Jones}, {Mandel}, {Muthukrishna}, {O'Grady},
  {Peters}, {Pierel}, {Ponder}, {Pr{\v s}a}, {Rodney}, \& {Villar}}]{kessler19}
{Kessler}, R., {Narayan}, G., {Avelino}, A., {et~al.} 2019, arXiv e-prints,
  arXiv:1903.11756

\bibitem[{{Kim} {et~al.}(2013){Kim}, {Thomas}, {Aldering}, {Antilogus},
  {Aragon}, {Bailey}, {Baltay}, {Bongard}, {Buton}, {Canto}, {Cellier-Holzem},
  {Childress}, {Chotard}, {Copin}, {Fakhouri}, {Gangler}, {Guy}, {Kerschhaggl},
  {Kowalski}, {Nordin}, {Nugent}, {Paech}, {Pain}, {Pecontal}, {Pereira},
  {Perlmutter}, {Rabinowitz}, {Rigault}, {Runge}, {Saunders}, {Scalzo},
  {Smadja}, {Tao}, {Weaver}, \& {Wu}}]{kim13}
{Kim}, A.~G., {Thomas}, R.~C., {Aldering}, G., {et~al.} 2013, \apj, 766, 84

\bibitem[{Kluyver {et~al.}(2016)Kluyver, Ragan-Kelley, P{\'e}rez, Granger,
  Bussonnier, Frederic, Kelley, Hamrick, Grout, Corlay, Ivanov, Avila, Abdalla,
  \& Willing}]{kluyver16}
Kluyver, T., Ragan-Kelley, B., P{\'e}rez, F., {et~al.} 2016, in Positioning and
  Power in Academic Publishing: Players, Agents and Agendas, ed. F.~Loizides \&
  B.~Schmidt, IOS Press, 87 -- 90

\bibitem[{{Knop} {et~al.}(2003){Knop}, {Aldering}, {Amanullah}, {Astier},
  {Blanc}, {Burns}, {Conley}, {Deustua}, {Doi}, {Ellis}, {Fabbro}, {Folatelli},
  {Fruchter}, {Garavini}, {Garmond}, {Garton}, {Gibbons}, {Goldhaber},
  {Goobar}, {Groom}, {Hardin}, {Hook}, {Howell}, {Kim}, {Lee}, {Lidman},
  {Mendez}, {Nobili}, {Nugent}, {Pain}, {Panagia}, {Pennypacker}, {Perlmutter},
  {Quimby}, {Raux}, {Regnault}, {Ruiz-Lapuente}, {Sainton}, {Schaefer},
  {Schahmaneche}, {Smith}, {Spadafora}, {Stanishev}, {Sullivan}, {Walton},
  {Wang}, {Wood-Vasey}, \& {Yasuda}}]{knop03}
{Knop}, R.~A., {Aldering}, G., {Amanullah}, R., {et~al.} 2003, \apj, 598, 102

\bibitem[{{Kowalski} {et~al.}(2008){Kowalski}, {Rubin}, {Aldering},
  {Agostinho}, {Amadon}, {Amanullah}, {Balland}, {Barbary}, {Blanc}, {Challis},
  {Conley}, {Connolly}, {Covarrubias}, {Dawson}, {Deustua}, {Ellis}, {Fabbro},
  {Fadeyev}, {Fan}, {Farris}, {Folatelli}, {Frye}, {Garavini}, {Gates},
  {Germany}, {Goldhaber}, {Goldman}, {Goobar}, {Groom}, {Haissinski}, {Hardin},
  {Hook}, {Kent}, {Kim}, {Knop}, {Lidman}, {Linder}, {Mendez}, {Meyers},
  {Miller}, {Moniez}, {Mour{\~a}o}, {Newberg}, {Nobili}, {Nugent}, {Pain},
  {Perdereau}, {Perlmutter}, {Phillips}, {Prasad}, {Quimby}, {Regnault},
  {Rich}, {Rubenstein}, {Ruiz-Lapuente}, {Santos}, {Schaefer}, {Schommer},
  {Smith}, {Soderberg}, {Spadafora}, {Strolger}, {Strovink}, {Suntzeff},
  {Suzuki}, {Thomas}, {Walton}, {Wang}, {Wood-Vasey}, \& {Yun}}]{kowalski08}
{Kowalski}, M., {Rubin}, D., {Aldering}, G., {et~al.} 2008, \apj, 686, 749

\bibitem[{{Kunz} {et~al.}(2007){Kunz}, {Bassett}, \& {Hlozek}}]{kunz07}
{Kunz}, M., {Bassett}, B.~A., \& {Hlozek}, R.~A. 2007, \prd, 75, 103508

\bibitem[{{Lochner} {et~al.}(2016){Lochner}, {McEwen}, {Peiris}, {Lahav}, \&
  {Winter}}]{lochner16}
{Lochner}, M., {McEwen}, J.~D., {Peiris}, H.~V., {Lahav}, O., \& {Winter},
  M.~K. 2016, \apjs, 225, 31

\bibitem[{{LSST Science Collaboration} {et~al.}(2009){LSST Science
  Collaboration}, {Abell}, {Allison}, {Anderson}, {Andrew}, {Angel}, {Armus},
  {Arnett}, {Asztalos}, {Axelrod}, {Bailey}, {Ballantyne}, {Bankert},
  {Barkhouse}, {Barr}, {Barrientos}, {Barth}, {Bartlett}, {Becker}, {Becla},
  {Beers}, {Bernstein}, {Biswas}, {Blanton}, {Bloom}, {Bochanski}, {Boeshaar},
  {Borne}, {Bradac}, {Brandt}, {Bridge}, {Brown}, {Brunner}, {Bullock},
  {Burgasser}, {Burge}, {Burke}, {Cargile}, {Chand rasekharan}, {Chartas},
  {Chesley}, {Chu}, {Cinabro}, {Claire}, {Claver}, {Clowe}, {Connolly}, {Cook},
  {Cooke}, {Cooray}, {Covey}, {Culliton}, {de Jong}, {de Vries}, {Debattista},
  {Delgado}, {Dell'Antonio}, {Dhital}, {Di Stefano}, {Dickinson}, {Dilday},
  {Djorgovski}, {Dobler}, {Donalek}, {Dubois-Felsmann}, {Durech},
  {Eliasdottir}, {Eracleous}, {Eyer}, {Falco}, {Fan}, {Fassnacht}, {Ferguson},
  {Fernandez}, {Fields}, {Finkbeiner}, {Figueroa}, {Fox}, {Francke}, {Frank},
  {Frieman}, {Fromenteau}, {Furqan}, {Galaz}, {Gal-Yam}, {Garnavich},
  {Gawiser}, {Geary}, {Gee}, {Gibson}, {Gilmore}, {Grace}, {Green}, {Gressler},
  {Grillmair}, {Habib}, {Haggerty}, {Hamuy}, {Harris}, {Hawley}, {Heavens},
  {Hebb}, {Henry}, {Hileman}, {Hilton}, {Hoadley}, {Holberg}, {Holman},
  {Howell}, {Infante}, {Ivezic}, {Jacoby}, {Jain}, {R}, {Jedicke}, {Jee},
  {Garrett Jernigan}, {Jha}, {Johnston}, {Jones}, {Juric}, {Kaasalainen},
  {Styliani}, {Kafka}, {Kahn}, {Kaib}, {Kalirai}, {Kantor}, {Kasliwal},
  {Keeton}, {Kessler}, {Knezevic}, {Kowalski}, {Krabbendam}, {Krughoff},
  {Kulkarni}, {Kuhlman}, {Lacy}, {Lepine}, {Liang}, {Lien}, {Lira}, {Long},
  {Lorenz}, {Lotz}, {Lupton}, {Lutz}, {Macri}, {Mahabal}, {Mandelbaum},
  {Marshall}, {May}, {McGehee}, {Meadows}, {Meert}, {Milani}, {Miller},
  {Miller}, {Mills}, {Minniti}, {Monet}, {Mukadam}, {Nakar}, {Neill}, {Newman},
  {Nikolaev}, {Nordby}, {O'Connor}, {Oguri}, {Oliver}, {Olivier}, {Olsen},
  {Olsen}, {Olszewski}, {Oluseyi}, {Padilla}, {Parker}, {Pepper}, {Peterson},
  {Petry}, {Pinto}, {Pizagno}, {Popescu}, {Prsa}, {Radcka}, {Raddick},
  {Rasmussen}, {Rau}, {Rho}, {Rhoads}, {Richards}, {Ridgway}, {Robertson},
  {Roskar}, {Saha}, {Sarajedini}, {Scannapieco}, {Schalk}, {Schindler},
  {Schmidt}, {Schmidt}, {Schneider}, {Schumacher}, {Scranton}, {Sebag},
  {Seppala}, {Shemmer}, {Simon}, {Sivertz}, {Smith}, {Allyn Smith}, {Smith},
  {Spitz}, {Stanford}, {Stassun}, {Strader}, {Strauss}, {Stubbs}, {Sweeney},
  {Szalay}, {Szkody}, {Takada}, {Thorman}, {Trilling}, {Trimble}, {Tyson}, {Van
  Berg}, {Vand en Berk}, {VanderPlas}, {Verde}, {Vrsnak}, {Walkowicz}, {Wand
  elt}, {Wang}, {Wang}, {Warner}, {Wechsler}, {West}, {Wiecha}, {Williams},
  {Willman}, {Wittman}, {Wolff}, {Wood-Vasey}, {Wozniak}, {Young}, {Zentner},
  \& {Zhan}}]{lsst09}
{LSST Science Collaboration}, {Abell}, P.~A., {Allison}, J., {et~al.} 2009,
  arXiv e-prints, arXiv:0912.0201

\bibitem[{{Malz} {et~al.}(2018){Malz}, {Hlo{\v z}ek}, {Allam}, {Bahmanyar},
  {Biswas}, {Dai}, {Galbany}, {Ishida}, {Jha}, {Jones}, {Kessler}, {Lochner},
  {Mahabal}, {Mandel}, {Mart{\'{\i}}nez-Galarza}, {McEwen}, {Muthukrishna},
  {Narayan}, {Peiris}, {Peters}, {Setzer}, {The LSST Dark Energy Science
  Collaboration}, {LSST Transients}, \& {Variable Stars Science
  Collaboration}}]{malz18}
{Malz}, A., {Hlo{\v z}ek}, R., {Allam}, Jr, T., {et~al.} 2018, arXiv e-prints,
  arXiv:1809.11145

\bibitem[{McKinney(2010)}]{mckinney10}
McKinney, W. 2010, in Proceedings of the 9th Python in Science Conference, ed.
  S.~van~der Walt \& J.~Millman, 51 -- 56

\bibitem[{{Oke} \& {Sandage}(1968)}]{oke68}
{Oke}, J.~B., \& {Sandage}, A. 1968, \apj, 154, 21

\bibitem[{{Okumura} {et~al.}(2014){Okumura}, {Ihara}, {Doi}, {Morokuma},
  {Pain}, {Totani}, {Barbary}, {Takanashi}, {Yasuda}, \&
  {Aldering}}]{okumura14}
{Okumura}, J.~E., {Ihara}, Y., {Doi}, M., {et~al.} 2014, \pasj, 66, 49

\bibitem[{{Pasquet} {et~al.}(2019){Pasquet}, {Pasquet}, {Chaumont}, \&
  {Fouchez}}]{pasquet19}
{Pasquet}, J., {Pasquet}, J., {Chaumont}, M., \& {Fouchez}, D. 2019, arXiv
  e-prints, arXiv:1901.01298

\bibitem[{Pedregosa {et~al.}(2011)Pedregosa, Varoquaux, Gramfort, Michel,
  Thirion, Grisel, Blondel, Prettenhofer, Weiss, Dubourg, Vanderplas, Passos,
  Cournapeau, Brucher, Perrot, \& Duchesnay}]{scikit-learn}
Pedregosa, F., Varoquaux, G., Gramfort, A., {et~al.} 2011, Journal of Machine
  Learning Research, 12, 2825

\bibitem[{{Perlmutter} {et~al.}(1999){Perlmutter}, {Aldering}, {Goldhaber},
  {Knop}, {Nugent}, {Castro}, {Deustua}, {Fabbro}, {Goobar}, {Groom}, {Hook},
  {Kim}, {Kim}, {Lee}, {Nunes}, {Pain}, {Pennypacker}, {Quimby}, {Lidman},
  {Ellis}, {Irwin}, {McMahon}, {Ruiz-Lapuente}, {Walton}, {Schaefer}, {Boyle},
  {Filippenko}, {Matheson}, {Fruchter}, {Panagia}, {Newberg}, {Couch}, \&
  {Project}}]{perlmutter99}
{Perlmutter}, S., {Aldering}, G., {Goldhaber}, G., {et~al.} 1999, \apj, 517,
  565

\bibitem[{{PLAsTiCC Team and PLAsTiCC Modelers}(2019)}]{plasticc_data}
{PLAsTiCC Team and PLAsTiCC Modelers}. 2019, {Unblinded Data for PLAsTiCC
  Classification Challenge},  Zenodo, doi:10.5281/zenodo.2539456.
\newblock \url{https://doi.org/10.5281/zenodo.2539456}

\bibitem[{{Poznanski} {et~al.}(2007){Poznanski}, {Maoz}, \&
  {Gal-Yam}}]{poznanski07}
{Poznanski}, D., {Maoz}, D., \& {Gal-Yam}, A. 2007, \aj, 134, 1285

\bibitem[{{Rasmussen} \& {Williams}(2006)}]{rasmussen06}
{Rasmussen}, C.~E., \& {Williams}, C.~K.~I. 2006, {Gaussian Processes for
  Machine Learning} (The MIT Press)

\bibitem[{{Revsbech} {et~al.}(2018){Revsbech}, {Trotta}, \& {van
  Dyk}}]{revsbech18}
{Revsbech}, E.~A., {Trotta}, R., \& {van Dyk}, D.~A. 2018, \mnras, 473, 3969

\bibitem[{{Richards} {et~al.}(2012){Richards}, {Homrighausen}, {Freeman},
  {Schafer}, \& {Poznanski}}]{richards12}
{Richards}, J.~W., {Homrighausen}, D., {Freeman}, P.~E., {Schafer}, C.~M., \&
  {Poznanski}, D. 2012, \mnras, 419, 1121

\bibitem[{{Riess} {et~al.}(1998){Riess}, {Filippenko}, {Challis},
  {Clocchiatti}, {Diercks}, {Garnavich}, {Gilliland}, {Hogan}, {Jha},
  {Kirshner}, {Leibundgut}, {Phillips}, {Reiss}, {Schmidt}, {Schommer},
  {Smith}, {Spyromilio}, {Stubbs}, {Suntzeff}, \& {Tonry}}]{riess98}
{Riess}, A.~G., {Filippenko}, A.~V., {Challis}, P., {et~al.} 1998, \aj, 116,
  1009

\bibitem[{{Riess} {et~al.}(2004){Riess}, {Strolger}, {Tonry}, {Casertano},
  {Ferguson}, {Mobasher}, {Challis}, {Filippenko}, {Jha}, {Li}, {Chornock},
  {Kirshner}, {Leibundgut}, {Dickinson}, {Livio}, {Giavalisco}, {Steidel},
  {Ben{\'{\i}}tez}, \& {Tsvetanov}}]{riess04}
{Riess}, A.~G., {Strolger}, L.-G., {Tonry}, J., {et~al.} 2004, \apj, 607, 665

\bibitem[{{Rigault} {et~al.}(2013){Rigault}, {Copin}, {Aldering}, {Antilogus},
  {Aragon}, {Bailey}, {Baltay}, {Bongard}, {Buton}, {Canto}, {Cellier-Holzem},
  {Childress}, {Chotard}, {Fakhouri}, {Feindt}, {Fleury}, {Gangler},
  {Greskovic}, {Guy}, {Kim}, {Kowalski}, {Lombardo}, {Nordin}, {Nugent},
  {Pain}, {P{\'e}contal}, {Pereira}, {Perlmutter}, {Rabinowitz}, {Runge},
  {Saunders}, {Scalzo}, {Smadja}, {Tao}, {Thomas}, \& {Weaver}}]{rigault13}
{Rigault}, M., {Copin}, Y., {Aldering}, G., {et~al.} 2013, \aap, 560, A66

\bibitem[{{Rodney} {et~al.}(2014){Rodney}, {Riess}, {Strolger}, {Dahlen},
  {Graur}, {Casertano}, {Dickinson}, {Ferguson}, {Garnavich}, {Hayden}, {Jha},
  {Jones}, {Kirshner}, {Koekemoer}, {McCully}, {Mobasher}, {Patel}, {Weiner},
  {Cenko}, {Clubb}, {Cooper}, {Filippenko}, {Frederiksen}, {Hjorth},
  {Leibundgut}, {Matheson}, {Nayyeri}, {Penner}, {Trump}, {Silverman}, {U},
  {Azalee Bostroem}, {Challis}, {Rajan}, {Wolff}, {Faber}, {Grogin}, \&
  {Kocevski}}]{rodney14}
{Rodney}, S.~A., {Riess}, A.~G., {Strolger}, L.-G., {et~al.} 2014, \aj, 148, 13

\bibitem[{{Rubin} {et~al.}(2015){Rubin}, {Aldering}, {Barbary}, {Boone},
  {Chappell}, {Currie}, {Deustua}, {Fagrelius}, {Fruchter}, {Hayden}, {Lidman},
  {Nordin}, {Perlmutter}, {Saunders}, {Sofiatti}, \& {Supernova Cosmology
  Project}}]{rubin15}
{Rubin}, D., {Aldering}, G., {Barbary}, K., {et~al.} 2015, \apj, 813, 137

\bibitem[{{Sako} {et~al.}(2011){Sako}, {Bassett}, {Connolly}, {Dilday},
  {Cambell}, {Frieman}, {Gladney}, {Kessler}, {Lampeitl}, {Marriner}, {Miquel},
  {Nichol}, {Schneider}, {Smith}, \& {Sollerman}}]{sako11}
{Sako}, M., {Bassett}, B., {Connolly}, B., {et~al.} 2011, \apj, 738, 162

\bibitem[{{Saunders} {et~al.}(2018){Saunders}, {Aldering}, {Antilogus},
  {Bailey}, {Baltay}, {Barbary}, {Baugh}, {Boone}, {Bongard}, {Buton}, {Chen},
  {Chotard}, {Copin}, {Dixon}, {Fagrelius}, {Fakhouri}, {Feindt}, {Fouchez},
  {Gangler}, {Hayden}, {Hillebrandt}, {Kim}, {Kowalski}, {K{\"u}sters},
  {Leget}, {Lombardo}, {Nordin}, {Pain}, {Pecontal}, {Pereira}, {Perlmutter},
  {Rabinowitz}, {Rigault}, {Rubin}, {Runge}, {Smadja}, {Sofiatti}, {Suzuki},
  {Tao}, {Taubenberger}, {Thomas}, {Vincenzi}, \& {Nearby Supernova
  Factory}}]{saunders18}
{Saunders}, C., {Aldering}, G., {Antilogus}, P., {et~al.} 2018, \apj, 869, 167

\bibitem[{{Scolnic} {et~al.}(2018){Scolnic}, {Jones}, {Rest}, {Pan},
  {Chornock}, {Foley}, {Huber}, {Kessler}, {Narayan}, {Riess}, {Rodney},
  {Berger}, {Brout}, {Challis}, {Drout}, {Finkbeiner}, {Lunnan}, {Kirshner},
  {Sanders}, {Schlafly}, {Smartt}, {Stubbs}, {Tonry}, {Wood-Vasey}, {Foley},
  {Hand}, {Johnson}, {Burgett}, {Chambers}, {Draper}, {Hodapp}, {Kaiser},
  {Kudritzki}, {Magnier}, {Metcalfe}, {Bresolin}, {Gall}, {Kotak}, {McCrum}, \&
  {Smith}}]{scolnic18}
{Scolnic}, D.~M., {Jones}, D.~O., {Rest}, A., {et~al.} 2018, \apj, 859, 101

\bibitem[{Stein(1999)}]{stein99}
Stein, M.~L. 1999, Interpolation of spatial data, Springer Series in Statistics
  (New York: Springer-Verlag), xviii+247, some theory for Kriging,
  doi:10.1007/978-1-4612-1494-6.
\newblock \url{http://dx.doi.org/10.1007/978-1-4612-1494-6}

\bibitem[{{Strolger} {et~al.}(2015){Strolger}, {Dahlen}, {Rodney}, {Graur},
  {Riess}, {McCully}, {Ravindranath}, {Mobasher}, \& {Shahady}}]{strolger15}
{Strolger}, L.-G., {Dahlen}, T., {Rodney}, S.~A., {et~al.} 2015, \apj, 813, 93

\bibitem[{{Suzuki} {et~al.}(2012){Suzuki}, {Rubin}, {Lidman}, {Aldering},
  {Amanullah}, {Barbary}, {Barrientos}, {Botyanszki}, {Brodwin}, {Connolly},
  {Dawson}, {Dey}, {Doi}, {Donahue}, {Deustua}, {Eisenhardt}, {Ellingson},
  {Faccioli}, {Fadeyev}, {Fakhouri}, {Fruchter}, {Gilbank}, {Gladders},
  {Goldhaber}, {Gonzalez}, {Goobar}, {Gude}, {Hattori}, {Hoekstra}, {Hsiao},
  {Huang}, {Ihara}, {Jee}, {Johnston}, {Kashikawa}, {Koester}, {Konishi},
  {Kowalski}, {Linder}, {Lubin}, {Melbourne}, {Meyers}, {Morokuma}, {Munshi},
  {Mullis}, {Oda}, {Panagia}, {Perlmutter}, {Postman}, {Pritchard}, {Rhodes},
  {Ripoche}, {Rosati}, {Schlegel}, {Spadafora}, {Stanford}, {Stanishev},
  {Stern}, {Strovink}, {Takanashi}, {Tokita}, {Wagner}, {Wang}, {Yasuda},
  {Yee}, \& {Supernova Cosmology Project}}]{suzuki12}
{Suzuki}, N., {Rubin}, D., {Lidman}, C., {et~al.} 2012, \apj, 746, 85

\bibitem[{{The Dark Energy Survey Collaboration}(2005)}]{des05}
{The Dark Energy Survey Collaboration}. 2005, arXiv e-prints, astro

\bibitem[{{The LSST Dark Energy Science Collaboration} {et~al.}(2018){The LSST
  Dark Energy Science Collaboration}, {Mandelbaum}, {Eifler}, {Hlo{\v z}ek},
  {Collett}, {Gawiser}, {Scolnic}, {Alonso}, {Awan}, {Biswas}, {Blazek},
  {Burchat}, {Chisari}, {Dell'Antonio}, {Digel}, {Frieman}, {Goldstein},
  {Hook}, {Ivezi{\'c}}, {Kahn}, {Kamath}, {Kirkby}, {Kitching}, {Krause},
  {Leget}, {Marshall}, {Meyers}, {Miyatake}, {Newman}, {Nichol}, {Rykoff},
  {Sanchez}, {Slosar}, {Sullivan}, \& {Troxel}}]{desc18}
{The LSST Dark Energy Science Collaboration}, {Mandelbaum}, R., {Eifler}, T.,
  {et~al.} 2018, arXiv e-prints, arXiv:1809.01669

\bibitem[{{van der Walt} {et~al.}(2011){van der Walt}, {Colbert}, \&
  {Varoquaux}}]{vanderwalt11}
{van der Walt}, S., {Colbert}, S.~C., \& {Varoquaux}, G. 2011, Computing in
  Science and Engineering, 13, 22

\bibitem[{{VanderPlas}(2018)}]{vanderplas18}
{VanderPlas}, J.~T. 2018, The Astrophysical Journal Supplement Series, 236, 16

\end{thebibliography}



\end{document}